\documentclass[12pt,a4paper]{article}
\pdfoutput=1
\usepackage{jcappub}
\usepackage{pstricks}
\usepackage{color}

\hypersetup{colorlinks,bookmarksopen,bookmarksnumbered,citecolor=blue,
linkcolor=black,pdfstartview=FitH,urlcolor=blue}

\graphicspath{ {figures/} }

\pdfoutput=1
\usepackage{graphicx}
\usepackage{amsmath}
\usepackage{hyperref}
\usepackage{amssymb}
\usepackage{epstopdf}
\usepackage{placeins}

\def\bea{\begin{eqnarray}}
\def\eea{\end{eqnarray}}
\def\beq{\begin{equation}}
\def\eeq{\end{equation}}

\newcommand{\lsim}{\mathrel{\rlap{\lower4pt\hbox{\hskip1pt$\sim$}}
    \raise1pt\hbox{$<$}}}         
\newcommand{\gsim}{\mathrel{\rlap{\lower4pt\hbox{\hskip1pt$\sim$}}
    \raise1pt\hbox{$>$}}}         

\newcommand{\leftrightarrowraised}{\mathrel{\rlap{\lower-0pt\hbox{\hskip1pt$\partial$}}
    \raise6 pt\hbox{$\leftrightarrow$}}}

\newcommand{\vect}[1]{\boldsymbol{\rm #1}}
\newcommand{\Ho}{H_{1\textrm{DM}}}
\newcommand{\Ht}{H_{2\textrm{DM}}}
\renewcommand{\L}{\mathcal{L}}
\newcommand{\T}{\mathcal{T}}

\renewcommand{\L}{\mathcal{L}}
\newcommand{\boldtheta}{\boldsymbol{\theta}}

\newcount\hour \newcount\minute
\hour=\time \divide \hour by 60
\minute=\time
\title{On the direct detection of multi-component dark matter: implications of the relic abundance}

\author[a,1]{Juan Herrero-Garcia,\note{\url{http://orcid.org/0000-0002-3300-0029}}}
\author[a,2]{Andre Scaffidi,\note{\url{http://orcid.org/0000-0002-1203-6452}}}
\author[a]{Martin White}
\author[a,3]{and Anthony G. Williams \note{\url{http://orcid.org/0000-0002-1472-1592}}}

\emailAdd{juan.herrero-garcia@coepp.org.au}
\emailAdd{andre.scaffidi@adelaide.edu.au}
\emailAdd{martin.white@adelaide.edu.au}
\emailAdd{anthony.williams@adelaide.edu.au}

\subheader{ADP-17-33/T1072}

\affiliation[a]{ARC Centre of Excellence for Particle Physics at the Terascale, Department of Physics, University of Adelaide, Adelaide, South Australia 5005, Australia}

\abstract{
Recently we studied the direct detection of multi-component dark matter with arbitrary local energy densities. Although the generation of the dark matter relic abundance is model-dependent, and in principle could be only indirectly related to direct detection, it is interesting to consider the implications of the former on the latter. In this work we conduct an extended analysis to include constraints from two natural scenarios of dark matter genesis: asymmetric dark matter and thermal freeze-out. In the first (second) case, the dark matter number (energy) densities of the different components are expected to be similar. In the case of thermal freeze-out, we assume that the global energy density scales with the local one. In our numerical analysis we analyse the median sensitivity of direct detection experiments to discriminate a two-component scenario from a one-component one, and also the precision with which dark matter parameters can be extracted. We analyse these generic scenarios for both light and heavy mediators. We find that most scenarios have a relatively suppressed maximum median sensitivity compared to the previously studied general cases. We also find that the asymmetric scenario is more promising than the thermal freeze-out one.}


\begin{document}
\maketitle

\section{Introduction} \label{sec:intro}

Some form of non-visible matter, termed dark matter (DM), constitutes a significant fraction of the universe. Several different particle physics scenarios have been proposed as an origin of this DM (see for instance Ref.~\cite{Bertone:2004pz} for a review). Among them, some of the best motivated ones are Weakly-Interacting-Massive-Particles (WIMPs), with their energy density originated from thermal freeze-out, and Asymmetric DM models (ADM)~\cite{Kaplan:2009ag}, where the abundance stems from an asymmetry in a similar fashion to the case of the observed baryon and lepton energy densities (see also Refs.~\cite{Zurek:2013wia,Petraki:2013wwa} for reviews on the topic). However, it is by no means guaranteed that just a single state or particle (1DM) constitutes the whole dark sector, which may have a multi-component nature as occurs in the visible sector.

In this work we focus on the direct detection (DD) of DM in underground detectors~\cite{Goodman:1984dc}. We extend our previous study of the DD of two-component (2DM) averaged rates~\cite{Herrero-Garcia:2017vrl}, in which we considered arbitrary local energy densities (see also Refs.~\cite{Profumo:2009tb,Batell:2009vb,Adulpravitchai:2011ei,Dienes:2012cf,Chialva:2012rq,Bhattacharya:2016ysw,Ahmed:2017dbb,Bhattacharya:2017fid}, and Ref.~\cite{Herrero-Garcia:2018lga} for the study of annual modulations), by taking into account the implications that reproducing the DM abundance has on DD. Under a given particle physics model, both the scattering and the annihilation cross section can be obtained, and as is well known, the global DM relic abundance can be obtained from the latter~\cite{Kolb:1990vq}. This approach was taken in Ref.~\cite{Profumo:2009tb}, where the authors related the WIMP-nucleon scattering cross-section to the annihilation one, motivated by supersymmetric models. By further assuming that the local energy density relevant for DD is proportional to the global one, one further reduces the allowed parameter space~\cite{Bertone:2010rv,Blennow:2015gta} (see also Ref.~\cite{Anderhalden:2012qt} for the validity of this assumption).

In this work we do not consider specific models, but concentrate on two generic particle physics scenarios: thermal freeze-out and ADM. In the asymmetric scenario the number densities of the two species are expected to be similar, whereas in the thermal case we assume that the local energy densities, which are of similar size for both components, scale as the global ones. For each scenario considered we study both heavy and light mediators. We test how well an average experiment can discriminate between the 1DM and the 2DM hypothesis as well as how accurately and precisely the DM parameters can be obtained from a positive signal.

The paper is structured as follows. In Sec.~\ref{sec:rate} we review the relevant notation for the DD of 2DM. In Sec.~\ref{sec:asymmetric} we study the case of ADM, while in Sec.~\ref{sec:thermal_fo} we consider the case of thermal WIMPs. Finally we give our main conclusions in Sec.~\ref{sec:conc}. We provide some details of the analysis methods used for hypothesis testing and parameter estimation in App.~\ref{app:analysis}. We give some expressions for thermal freeze-out scenarios with light mediators in App.~\ref{app:loc_glob_simple}.
 
\section{Direct detection of two-component dark matter}  \label{sec:rate}
We present in the following the relevant expressions for the DD of 2DM scenarios. See Ref.~\cite{Herrero-Garcia:2017vrl} for more details and expressions for the general case of multiple components. 

\subsection{The \emph{general} scenario} 
We take the DM components to have masses $m_1<m_2$, spin-independent (SI) and isospin-conserving cross-sections with protons $\sigma_1^p,\,\sigma_2^p$, and local energy densities $\rho_1,\,\rho_2$, such that $\rho_1+\rho_2=\rho_{\rm loc}$, which we fix to be $0.4 \, {\rm GeV/cm}^3$. We define
\beq\label{eq:rhos} 
r_{\rho} \equiv \frac{\rho_2}{\rho_1}\,,\qquad  r_{\sigma} \equiv \frac{\sigma_2^p}{\sigma_1^p}\,,
\eeq 
such that $\rho_2 = r_\rho\,\rho_{\rm loc}/(1+r_\rho)$. We can write the averaged rate over the year, typically measured in events/(kg keV day), for a detector with target nucleus labelled by $(A,Z)$ (with mass $m_A$), as
\begin{align}
\label{eq:rate_tot}
R_{A}(E_{R}) &=R_{A}^{(1)}(E_R) + R_{A}^{(2)}(E_R) \nonumber\\&=  \frac{\rho_{\rm loc}\,\sigma_1^p}{2 \,(1+r_\rho)\,\mu_{p1}^2\,m_1} \,A^2\,F_{A}^2(E_{R}) \,\left[\eta(v_{m,A}^{(1)})+\,r_{\rho}\,r_{\sigma}\, \frac{\mu_{p1}^2\,m_1}{\mu_{p2}^2\,m_2}\,\eta(v_{m,A}^{(2)})\right]\,.
\end{align}
Here $F_A(E_{R})$ is the SI nuclear form factor of element with mass number $A$, for which we use the Helm parametrisation~\cite{PhysRev.104.1466,LEWIN199687}, and $\mu^2_{p}$ is the DM-proton reduced mass. In addition to $\rho_{\beta}$ ($\beta=1,2$ for the two DM particles), the astrophysics enters in Eq.~\eqref{eq:rate_tot} through the halo integral
\beq\label{eq:eta} 
\eta (v_{m,A}^{(\beta)}) =
\int_{v > v_{m,A}^{(\beta)}} \negthickspace \negthickspace d^3 v \frac{f^{(\beta)}_{\rm det}(\vect{v})}{v}, \, \qquad \text{with}\qquad v^{(\beta)}_{m,A}(E_{R})=\sqrt{ \frac{m_A E_{R}}{2 \mu_{\beta A}^2}}\,,
\eeq
where $v_{m,A}^{(\beta)}(E_{R})$ is the minimum velocity of the DM particle $\beta=1,2$ required to produce a nuclear recoil of energy $E_R$, and $\mu_{\beta A}$ is the DM-nucleus reduced mass.
$f_{\rm det}(\vect v, t)$ describes the distribution of DM particle velocities in the detector rest-frame, which we take to have the same functional form for both species. It can be written in terms of the galactic velocity distributions by doing a Galilean boost, $f_{\rm det}(\vect{v},t) = f_{\rm gal}(\vect{v} + \vect{v}_e)$, where $\vect{v}_e$ is the velocity vector of the Earth in the galactic rest-frame.\footnote{We do not include the effects of the gravitational focusing (GF)  by the Sun (see Refs.~\cite{Alenazi:2006wu,Lee:2013wza,Bozorgnia:2014dqa}) since it only affects the phases of the annual modulation and hence has no bearing on the average rates considered in this work. We include the effects of GF in Ref.~\cite{Herrero-Garcia:2018lga} in the context of annual modulations, where it is a relevant effect.} As we are considering only averaged rates, we use a constant $|\mathbf{v_e}|=230 \,\rm km\, s^{-1}$. We use a Maxwellian velocity distribution $f_{\rm gal}(v)=\frac{1}{(2 \pi {\sigma}^2_{H_\beta})^{3/2}} \exp{\Big(-\frac{3v^2}{2{\sigma}_{H_\beta}^2}\Big)}$, with velocity dispersion ${\sigma}_{H_\beta}$ and a cut-off at the escape velocity $v_{\rm esc} =550\,\rm km\, s^{-1}$. Notice that the velocity dispersions of both components can be different, as we discuss in the following section. We refer to Eq.~\eqref{eq:rate_tot} as the \textit{general} scenario.\footnote{In the following whenever the \textit{general} scenario is plotted we fix $r_\rho=r_\sigma=1$.} 

Apart from an overall normalisation, the slopes of the energy spectra of the different DM components are governed primarily by the term in squared brackets (and to a lesser extent the form factor). As discussed in Ref.~\cite{Herrero-Garcia:2017vrl}, the smoking gun signature of 2DM is the presence of a \emph{kink} in the total rate. This is because (in the \emph{general} model at least) the light component generates a spectrum that falls off more rapidly at low $E_R$ than the heavy component. This distinctive feature rapidly vanishes for smaller mass splittings, when both components have the same slope, and also for very large values of the heavy DM mass, for which the number density of the heavy component is very suppressed. Notice in Eq.~\eqref{eq:rate_tot} that the DM masses (and the form factor) control the slopes of the spectra, while the ratio of the rates of the heavy and light components at a certain energy, $R_A^{(2)} (E_R)/R_A^{(1)} (E_R)$, specifies the position of the \emph{kink}. We define $r_R$ to be the factor in front of the $\eta(v_{m,A}^{(2)})$ term, i.e., $R_A^{(2)}/R_A^{(1)}\equiv r_R\, \eta^{(2)}/\eta^{(1)}$, which hence summarises the different underlying particle physics of the scenarios considered. In the third column of Tab.~\ref{tab:benchmarks} we present $r_R$ for each scenario considered in this study.

As before we have checked that sensible energy resolutions do not significantly affect the detection rates, and in the following analysis we assume perfect energy resolution and efficiency. Hence, our results are intended to understand the underlying physics and more sophisticated experimental studies are needed once a signal is observed. In the numerical analysis we consider xenon and germanium targets, as they have quite different masses. For xenon (germanium) we use an exposure of 3\,(4) $t \cdot y$ with an energy threshold of $1\,(2)$ keV. We note that the median sensitivity scales with $\sim\sqrt{\rm exposure}$ (before systematics dominate) and hence our results can be rescaled for other exposures.
\begin{table}[t!]	
  	\centering{
  		\begin{tabular}{cccccccc}
  			Scenario & Mediator & $r_R$  & $m_1$ (GeV) &$m_2$ (GeV) & $r_\rho$ &  $r_\sigma$   & $m_{A^\prime} $ (MeV) \\ \hline
  			Asymmetric  & Heavy & $r_{\sigma}\, \mu_{p1}^2/\mu_{p2}^2$&10 & 50& -&1& - \\\hline
  			Asymmetric &Light & 1& 10 & 50& -&-& $30$ \\\hline
  			Freeze-out  & Heavy &$m_1^3/m_2^3$& 10 & 40& 10$^{-3}$&-& - \\\hline
  			Freeze-out &  Light & $m_2/m_1$& 8 & 20& 1& -& 1 \\\hline
  		\end{tabular}}
  		\caption{Benchmark points used to generate the Asimov data for parameter estimation. Cases where certain parameters have been absorbed into other degrees of freedom or are not applicable are shown with a dash. All coupling strengths are fixed to either $\sigma^1_p=10^{-45}$ cm$^2$ or $\epsilon_\text{eff}=\alpha\,\epsilon^2=10^{-22}$. Also shown in the third column is the factor $r_R$ that determines the position of the kink in the rate of 2DM scenarios.}
  		\label{tab:benchmarks}
  	\end{table}

\begin{figure}[t!]
	\centering
	\includegraphics[scale=0.3]{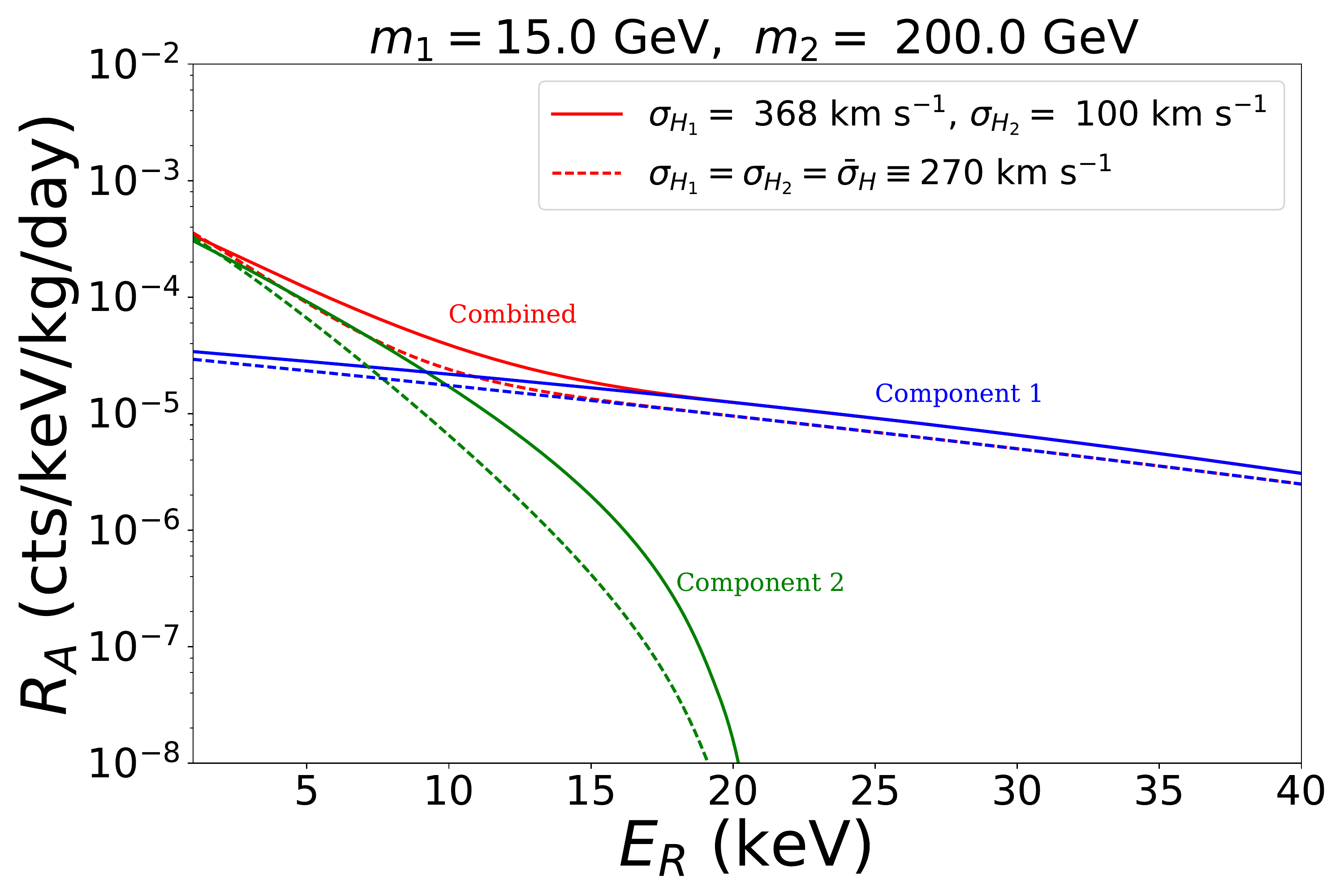}
	\caption{Differential event rates in xenon for mass-dependent velocity dispersions according to Eq.~\eqref{eq:MD_sigma} (solid), and for constant ones equal to $ \bar{\sigma}_{H}= 270$ kms$^{-1}$ (dashed), for $m_1=15$ GeV (green), $m_2=200$ GeV (blue) and the combined rates (red).
	}	
	\label{fig::vel_disp_comparrison}
\end{figure}

\subsection{Velocity dispersions}  \label{subsec:vel_disp}
The effective temperature for an isometric Maxwellian thermal distribution is given by $T_\beta \simeq \frac{1}{2}\,m_\beta\, \bar{\sigma}_H^2$, where we take a canonical value for the velocity dispersion, $ \bar{\sigma}_{H}\sim 270$ kms$^{-1}$. This is valid for the single species case, and for multiple species $N$ (labelled by $\beta=1,\,2...,\,N$) that interact with each other. For multiple species which are self-interacting on time scales much shorter than those relevant for the halo formation, the dark plasma attains hydrostatic equilibrium at a temperature $T_{\rm eq}$ on galactic scales. The result is that the Maxwellian velocity dispersion of each component of the dark sector, ${\sigma}_{H_\beta}$, becomes mass dependent~\cite{Foot:2012cs},
\begin{align} \label{eq:MD_sigma}
{\sigma}_{H_\beta} = \bar{\sigma}_H\sqrt{\frac{\bar{m}}{m_\beta}}\;,
\end{align}        
where $\bar{m} = \sum\limits_\beta n_\beta m_\beta/\sum\limits_\beta n_\beta$, with $n_\beta=\rho_\beta/m_\beta$ the number density of the DM particle $\beta$. As can be seen, the heavier DM particle has a narrower velocity distribution than the lighter one.

In Fig.~\ref{fig::vel_disp_comparrison} we show how the event rate changes between equal velocity dispersions (dashed curves) equal to $\bar{\sigma}_{H}$, and the mass-dependent case (solid curves). We use $m_1=15$ GeV (red), $m_2=200$ GeV (blue), and fixed $r_\rho=1$. In this case, ${\sigma}_{H_1}= 368>\bar{\sigma}_H$ and  ${\sigma}_{H_2}=100<\bar{\sigma}_H$. We observe that there are no large differences between both cases, although the rates slightly increase in the case of mass-dependent velocity dispersions. However, in this case the \emph{kink} of the 2DM spectrum is more smooth than for constant velocity dispersions, and therefore will be harder to observe.\footnote{We refer the reader to our previous study in Ref.~\cite{Herrero-Garcia:2017vrl} where we show that 2DM spectra that have less pronounced kinks are statistically less significant to discriminate from the 1DM case.}

In the following we consider 2DM scenarios that have either heavy or light mediators. For the case of a light mediator DM self-interactions are not suppressed, and therefore it seems more natural that the DM components have mass-dependent velocity dispersions. However, as the difference with the case of constant velocity dispersion is not very large, in order to more readily investigate the differences between the different cases, in the following we take constant velocity dispersions equal to $\bar{\sigma}_{H}$ for each DM component.  

\section{Asymmetric dark matter} \label{sec:asymmetric}
The production mechanism of DM is unknown. One of the most natural ways to obtain the correct dark matter relic abundance is via an asymmetry, in the same spirit as is expected to happen in the baryon and lepton sectors \cite{Petraki:2013wwa,Foot:2012cs}. ADM scenarios also naturally feature multi-component DM for models with a light or massless force carrier which follows from the requirement of gauge symmetry and the possible formation of bound states~\cite{Baldes:2017gzu}. 

\subsection{Heavy mediator}  \label{subsec:equal_n}
In this first case, we assume that there is an effective point-like interaction generated by some heavy new physics. We also assume that the two DM particles have similar number densities, $n_1=n_2$, so that
 \begin{align}
r_\rho = \frac{m_2}{m_1}\,,
 \end{align}
and the event rate in Eq.~\eqref{eq:rate_tot} becomes 
 \begin{align}
 \label{eq:rate_tot_neqn}
 R_{A}(E_{R}) &=  \frac{\rho_{\rm loc}\,\sigma_1^p }{2 \,(m_1+m_2)\,\mu_{p1}^2} \,A^2\,F_{A}^2(E_{R}) \,\left[\eta(v_{m,A}^{(1)}) +\,r_{\sigma}\, \frac{\mu_{p1}^2}{\mu_{p2}^2}\,\eta(v_{m,A}^{(2)})\right]\,.
 \end{align}
The simplification in this expression is due to the fact that the individual rates are proportional to the (equal) DM number densities. In most asymmetric models the number densities of DM and baryons are related by order one factors, and therefore one expects DM masses $\mathcal{O}(5)$ GeV to reproduce the DM relic abundance. This is well-motivated but model-dependent, and in the following we consider ADM-like scenarios with DM masses of up to a few hundreds GeV.

\begin{figure}[!t]
 	\centering
 	\includegraphics[scale=0.26]{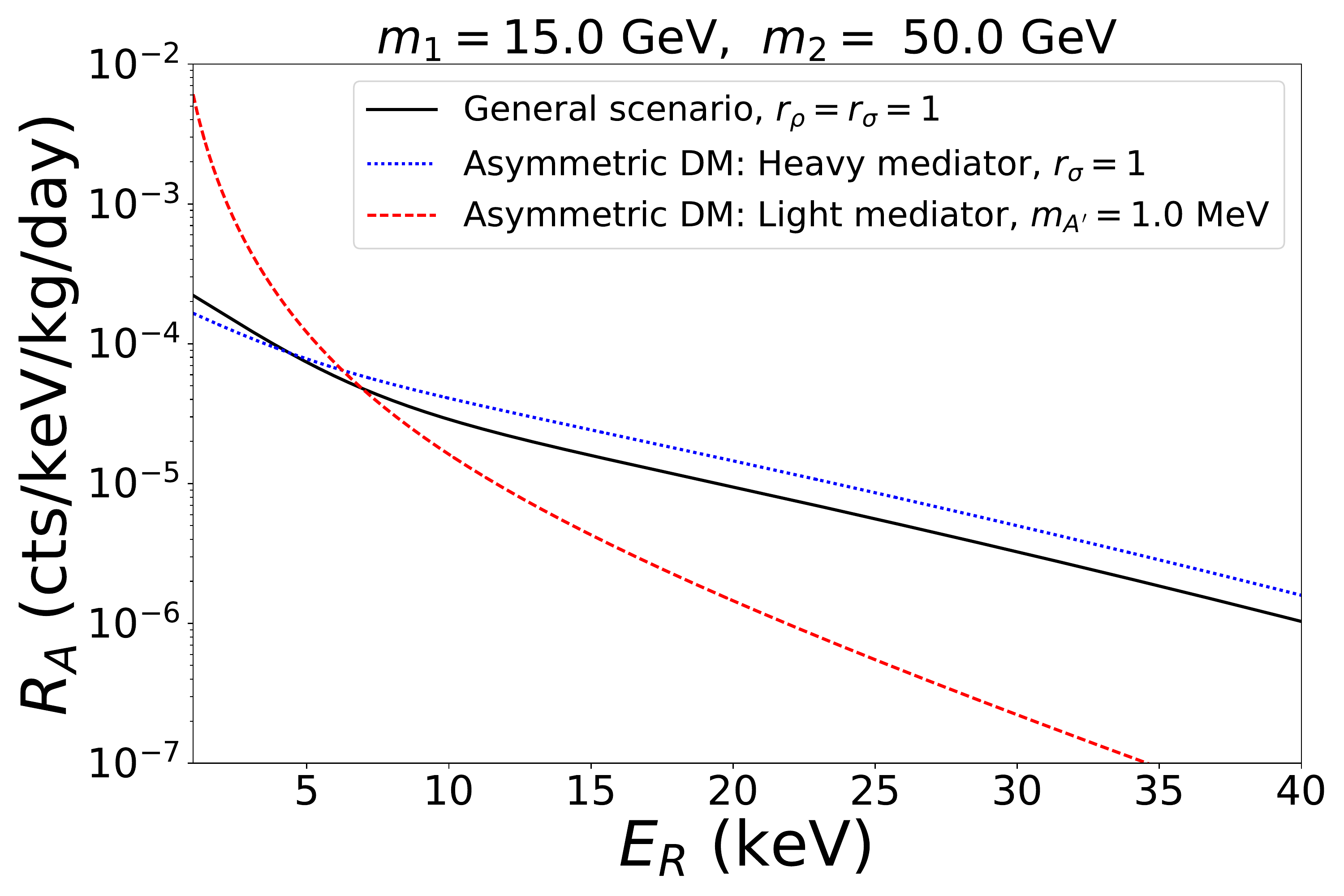}~~\includegraphics[scale=0.26]{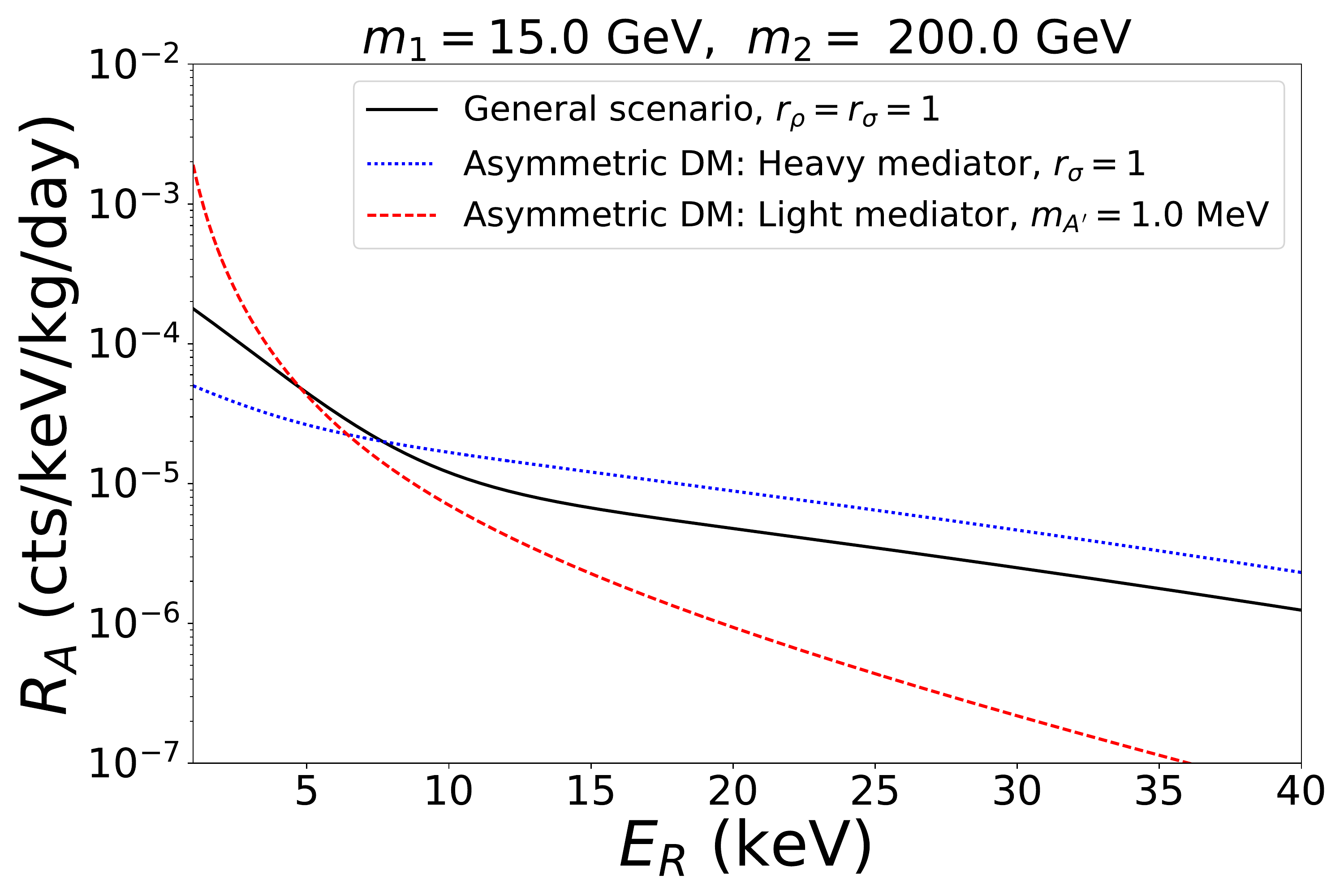}
 	\caption{Differential event rates in a xenon experiment for the \emph{general} scenario (solid black), for equal number densities with a heavy mediator (dotted blue), and for the case of light mediators with $m_{A^\prime} = 1$ MeV (dashed red). We fix $m_1=15$ GeV. \emph{Left:} $m_2=50$ GeV. \emph{Right:} $m_2=200$ GeV.}
 	\label{fig::compare_old_scan_spectra}
 \end{figure}
 
In Fig.~\ref{fig::compare_old_scan_spectra} we show the detection rates for the \emph{general} scenario, i.e.,  Eq.~\eqref{eq:rate_tot} (solid black), for equal number densities with a heavy mediator (dotted blue), and for the case of light mediators with $m_{A^\prime} = 1$ MeV (dashed red). The only quantitative difference between the spectra of the equal number density case and the \emph{general} scenario (studied in detail in Ref.~\cite{Herrero-Garcia:2017vrl}) is the fact that now we are fixing $r_\rho=m_2/m_1$, which leads to a smoothened \emph{kink} feature in the recoil spectrum. The result is that the resolving power of the 1DM versus 2DM hypothesis is expected to be worse for ADM-like models.

 \begin{figure}[h]
 	\centering
 	\includegraphics[scale=0.35]{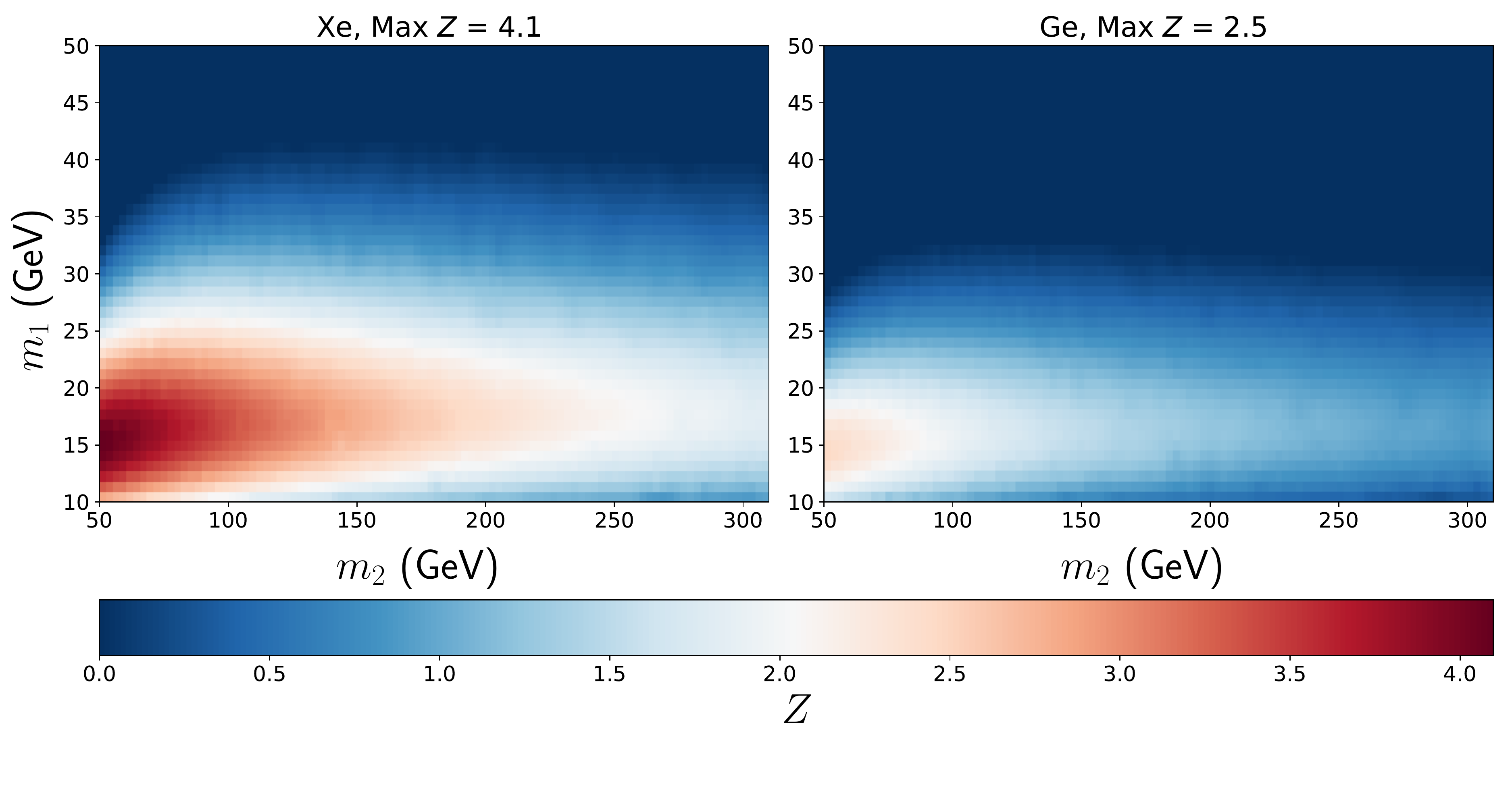}\\
 	\includegraphics[scale=0.35]{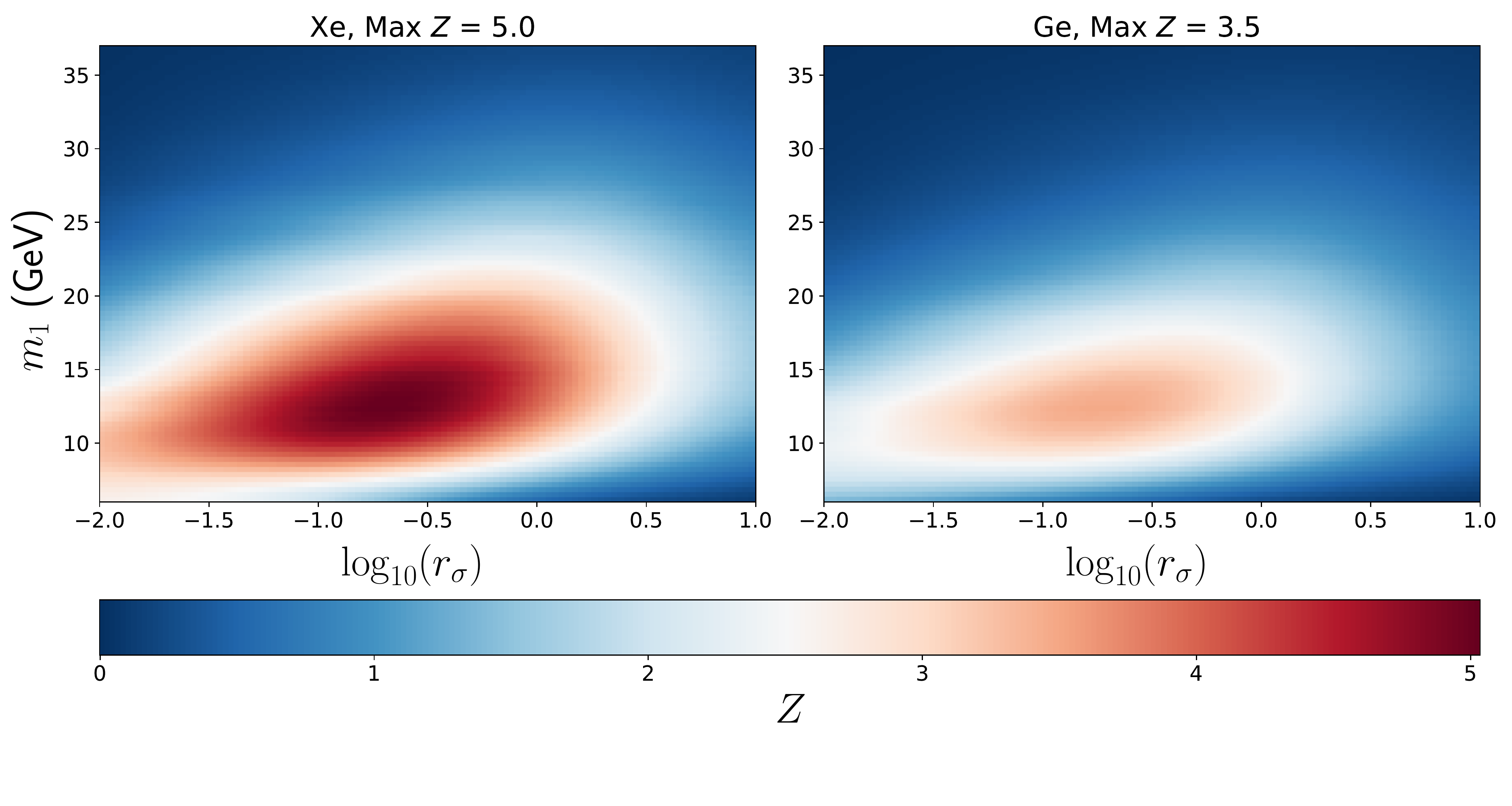}
 	\caption{Median significance $Z$  with which an ADM scenario with heavy mediators  can reject the 1DM hypothesis in favour of the 2DM hypothesis. \emph{Top:} $m_2-m_1$ plane, with fixed $r_\sigma= 1$. \emph{Bottom:} $\log_{10}(r_\sigma)-m_1$ plane, with fixed $m_2=50$ GeV. The left panel is for Xe, and the right one for Ge. }
 	\label{fig::eq_num_des_m1m2}
 \end{figure}
 
To quantify this, we plot in Fig.~\ref{fig::eq_num_des_m1m2} the `median significance'  for such discrimination in the case of ADM with a heavy mediator, and compare it to our previous study, namely Fig. 7 of Ref.~\cite{Herrero-Garcia:2017vrl} for xenon and germanium type detectors\footnote{Note we do not include sodium as we did in our previous study since the median significance was always found to be heavily suppressed relative to the Xe and Ge cases.}. For details and definitions of the analysis used to forecast the median sensitivity of an experiment to discriminating between 1DM and 2DM see App.~\ref{app:hypothesis}. In Fig.~\ref{fig::eq_num_des_m1m2} we show the results of the hypothesis testing in the planes $m_2-m_1$ (top) and $\log_{10}(r_{\sigma})-m_1$ (bottom). The left panel is for Xe and the right one for Ge. We see that the maximum median significance achieved for the heavy mediator model (max $Z\sim5$ for Xe) relative to the \emph{general} case (max $Z\sim 12.8$ for Xe) is reduced. We notice throughout this work that xenon-type experiments typically give a better median significance overall than germanium type experiments, with the exception of thermal freeze-out scenarios with light mediators (Sec.~\ref{sec::fo_light_med}). Aside from the overall reduction in median sensitivity,  as can be seen in the top panel of Fig.~\ref{fig::eq_num_des_m1m2}, the point of maximum $Z$ is shifted to lower $m_2$ in the heavy mediator model. This is because at low $m_2$ the total rate gets larger at low recoil energies, producing more events and hence increasing $Z$. Furthermore, as can be seen in the lower panel of Fig.~\ref{fig::eq_num_des_m1m2}, there is an upper cut-off in the lobe at $r_\sigma\sim 3$. This is due to the enhancement in the $r_R$ factor at large $r_\sigma$ that drives the rate to look like a single heavy component.

\begin{figure}[h!]
	\centering
	\includegraphics[scale=0.44]{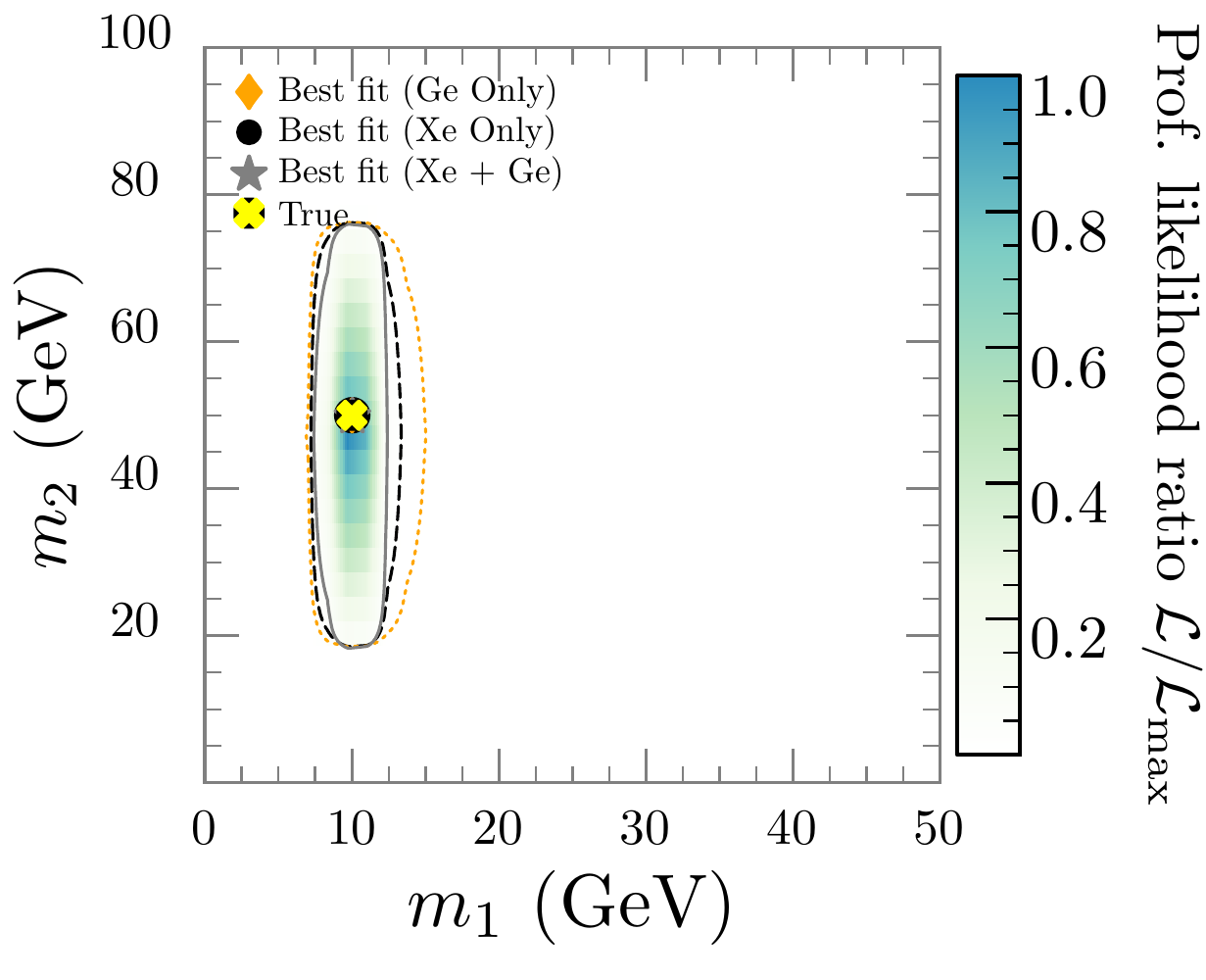}~
	\includegraphics[scale=0.44]{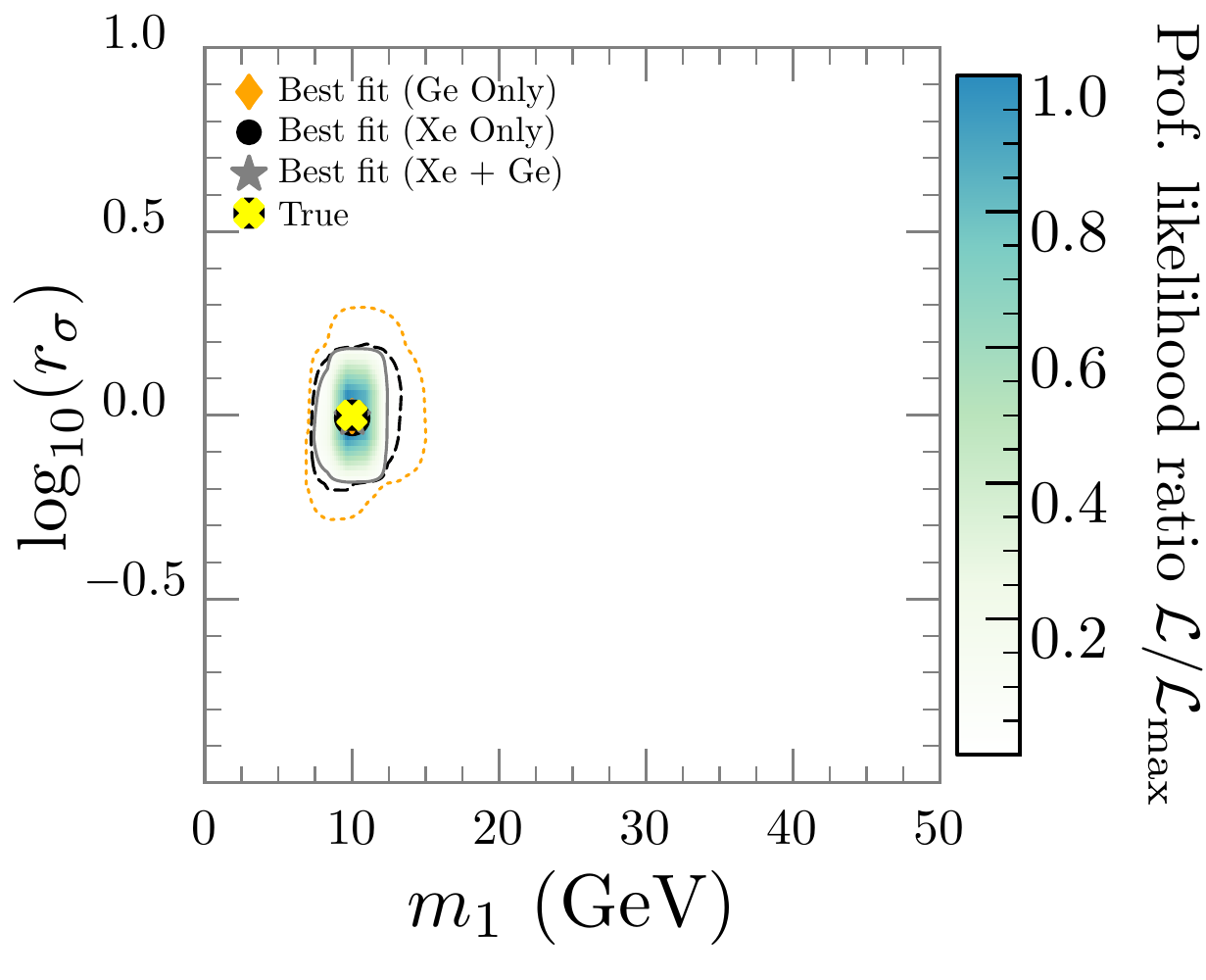}~
	\includegraphics[scale=0.44]{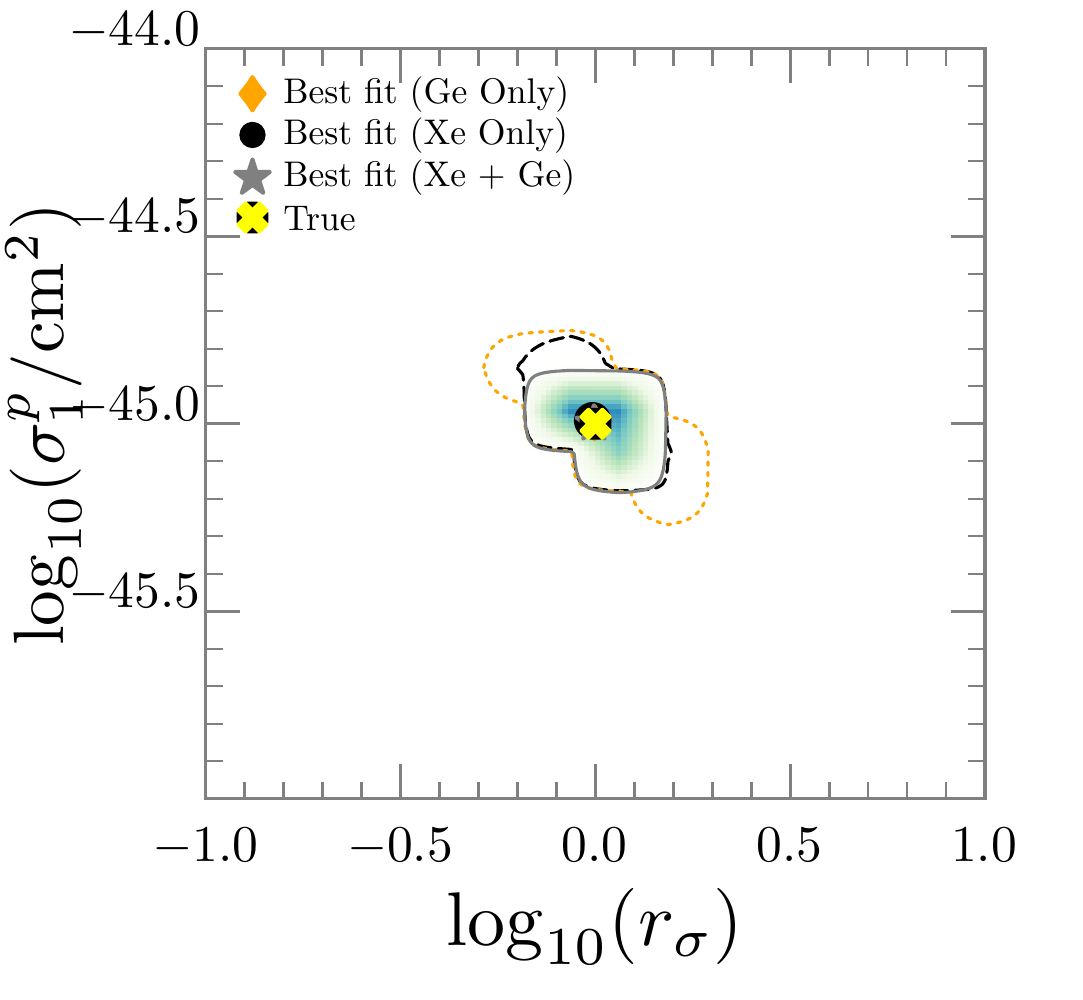}
	\caption{PLR density for the ADM scenario with  a heavy mediator. \emph{Left: }$m_1-m_2$ plane.  \emph{Middle: } $m_1-\log_{10}(r_\sigma)$ plane.   \emph{Right: } $\log_{10}(r_\sigma)-\log_{10}(\sigma_1^p/\text{cm}^2)$ plane. Contours represent $2\sigma$ C.L regions.   Benchmark points for generating the Asimov data are given in Tab.~\ref{tab:benchmarks}. The different colour contours correspond to a xenon only (black dashed), germanium only (orange dotted) and combined xenon + germanium (grey solid) experiments.}
	\label{fig:eq_num_param_est}
\end{figure}

Next we show in Fig.~\ref{fig:eq_num_param_est} how well the parameters can be estimated in the case of  a heavy mediator by calculating the Asimov likelihood and generating the Profile Likelihood Ratio (PLR) in 2D regions of the parameter space. We use the analysis methods summarised in App.~\ref{app:param_est}. The benchmarks used in this study to generate the Asimov data are provided in Tab.~\ref{tab:benchmarks}. These benchmark values belong to regions of parameter space which give a large discrimination between the 1DM and the 2DM hypotheses as ascertained by the results of the hypothesis testing. We show $2\sigma$ C.L regions for the first benchmark point given in Tab.~\ref{tab:benchmarks} in the $m_1-m_2$ plane (left panel), $m_{1}(\text{GeV})-\log_{10}(r_\rho)$ plane (middle panel) and the $\log_{10}(r_\rho)-\log_{10}(\sigma_1^p)$ plane. The different colour contours correspond to a xenon only (black dashed), germanium only (orange dotted) and xenon + germanium combined (grey solid) experiment.  We observe no degeneracies in the PLR with excellent parameter reconstruction.

 \subsection{Light mediators} \label{subsec:light_med}
We now consider the case that the two species are produced in some asymmetric scenario  but are couple to a light vector mediator with mass $m_{A^\prime}$. We consider an extra $U(1)$ gauge symmetry with dark coupling $g_D$, such that the gauge theory becomes $SU(3)_C\times SU(2)_L\times U(1)_Y\times U(1)_D$. As the dark symmetry is Abelian, a Stueckelberg mass for the dark photon is gauge invariant, but it can also be generated by a new scalar that takes a VEV.  The fermionic DM component $\beta$ ($\beta=1,2$) has charge $Q_{\beta}$ under this new $U(1)_D$ symmetry. For definiteness we take $Q_2=1$. We consider the case where there is an induced kinetic mixing $\epsilon$ between the dark and the visible photons parameterised by a term in the Lagrangian  $-\frac{\epsilon}{2}F_{\mu\nu}F^\prime_{\mu\nu}$. For our purposes, it is sufficient to consider the interaction terms generated at low energies after diagonalising the gauge boson kinetic terms,
\begin{align}
\label{intLagKin}
\L_\text{int} \supset \epsilon\,e J^\mu_{\rm EM}{A^\prime}_\mu +g_D\, J^\mu_{\rm DM}{A^\prime}_\mu \;,
\end{align}   
where $e$ is the magnitude of the electric charge, $J^\mu_{\rm EM} \equiv \sum_f q_f \bar{f}\gamma^\mu f$ is the electromagnetic current of Standard Model fermion $f$ (with electric charges $q_f$ in units of e) and $J_{\rm DM}^\mu \equiv \sum_{\beta}Q_\beta\bar{\chi_\beta}\gamma^\mu \chi_\beta$ is the dark fermion current. These are the interaction terms that are relevant for tree-level DD. The scattering cross section of DM component $\beta$ reads
\begin{align}
\label{eq::sigmap}
\bar \sigma_\beta^p(|\vec q|) = \sigma_\beta^p \times \frac{m_{A^\prime}^4}{(|\vec q|^2+m_{A^\prime}^2)^2}\,,
\end{align} 
where $\vec q$ is the momentum transfer, whose absolute value can be expressed in terms of the recoil energy as $|\vec q|=\sqrt{2m_AE_R}$, and $\sigma^p_\beta$ is the zero-momentum WIMP-proton cross-section of DM $\beta$, given by 
\begin{align}
\label{eq::sigma0}
\sigma_\beta^p = \frac{16\pi\,\epsilon_\text{eff}\,\alpha_\text{EM}\,\,\mu_{p_\beta}^2\:Q^2_\beta}{m_{A^\prime}^4}\,,
\end{align} 
where $\epsilon_\text{eff}\equiv \epsilon^2\alpha_{\text{D}}$ and  $\alpha_\text{EM}=e^2/(4\pi)$ [$\alpha_D=g^2_D/(4\pi)$] is the electromagnetic [dark] fine structure constant. The universal neutrality of the $U(1)_D$ charge implies that a plasma containing two species has charges of opposite sign with 
\begin{align}
\label{eq::charge_con}
n_1Q_1 + n_2 Q_2 = 0\;.
\end{align} 
As we assume that the species have equal number density $n_1=n_2$, this implies that $Q_2 = -Q_1 =1$, and hence $r_\sigma = \mu^2_{p_2}/\mu^2_{p_1}$. With this in mind, the total DD rate is simplified to
\begin{align}
\label{eq::light_med_rate}
 R_A(E_{R}) &=  \frac{\rho_{\rm loc}\,\sigma_1^p \,{Z^2} }{2 \,(m_1+m_2)\,\mu^2_{p_1}} \,F_{A}^2(E_{R})\,\left[\,\eta(v_{m,A}^{(1)}) +\eta(v_{m,A}^{(2)})\right]\times \frac{m_{A^\prime}^4}{(2m_AE_R+m_{A^\prime}^2)^2}\;.
\end{align}
One should note that since kinetic mixing only produces interactions with the SM photons, the DM couples to the electric charge. Therefore, the rate is proportional to the number of protons $Z$, and not to $A$. 

In Fig.~\ref{fig::compare_old_scan_spectra} we plot the rate of Eq.~\eqref{eq::light_med_rate} for $m_{A^\prime} = 1$ MeV as a red dashed curve. The momentum-dependent suppression factor $1/(q^2+m_{A^\prime}^2)$ in the scattering cross-section, Eq.~\eqref{eq::sigmap}, produces a steeper decrease with energy of the event rate. One should immediately notice that whilst there is a slight enhancement in the rate at low recoil energies, there is an absence of any distinct \emph{kink} feature in the spectrum. We therefore expect the maximum median sensitivity $Z$ in the case of an asymmetric scenario with a light mediator to be small except for large values of the effective mixing $\epsilon_\text{eff}$ or very small mediator masses $m_{A^\prime}$ (compared to $\vec q$), both of which give enhancements in $\sigma_1^p(|\vec q|)$.\footnote{Also notice that in the limit of large $m_{A^\prime}\gg |\vec q|$ we recover the contact case, where the discrimination is larger than for light mediators.}

 \begin{figure}[h!]
 	\centering
 	\includegraphics[scale=0.32]{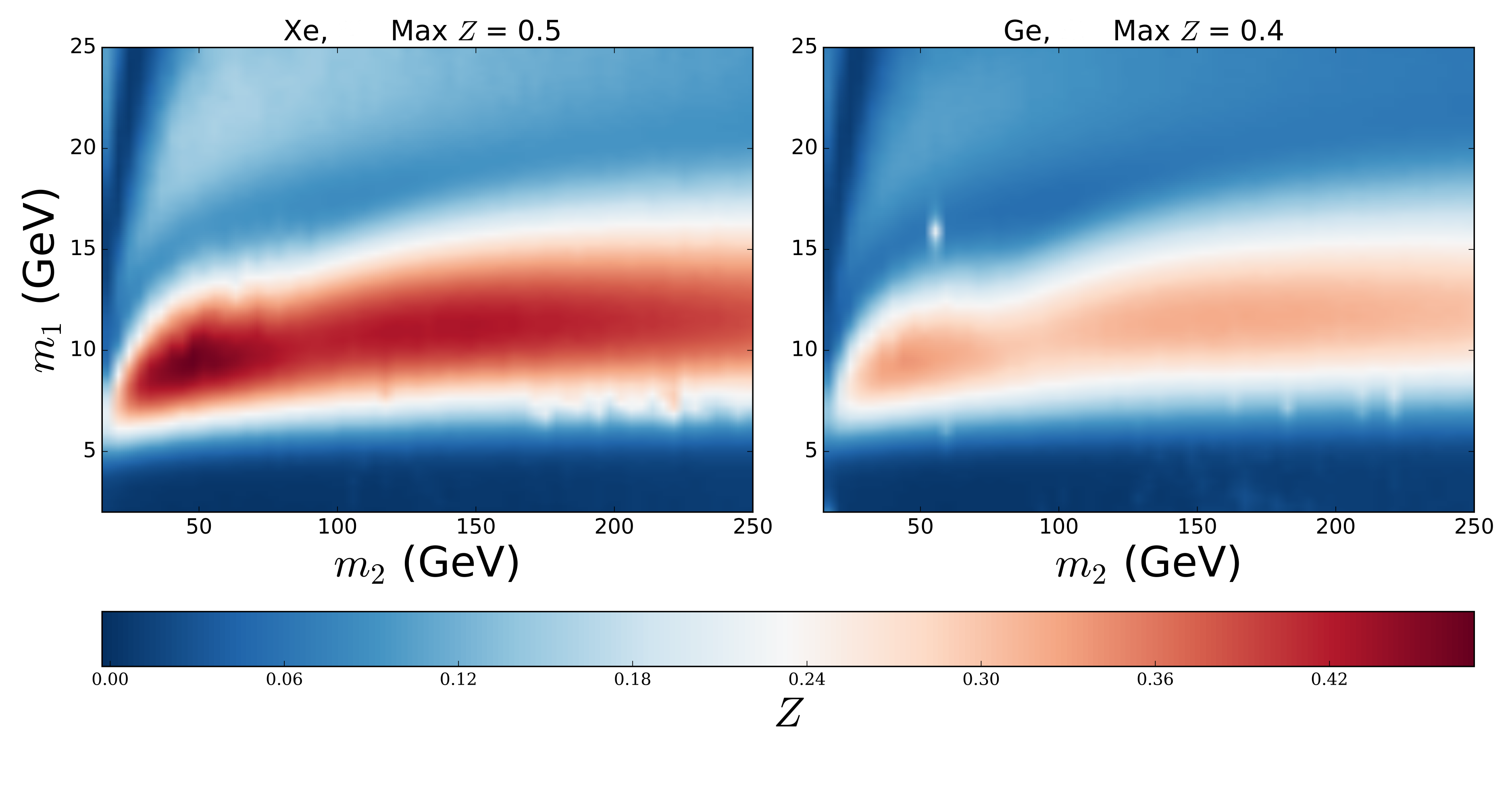}\\
 	\includegraphics[scale=0.32]{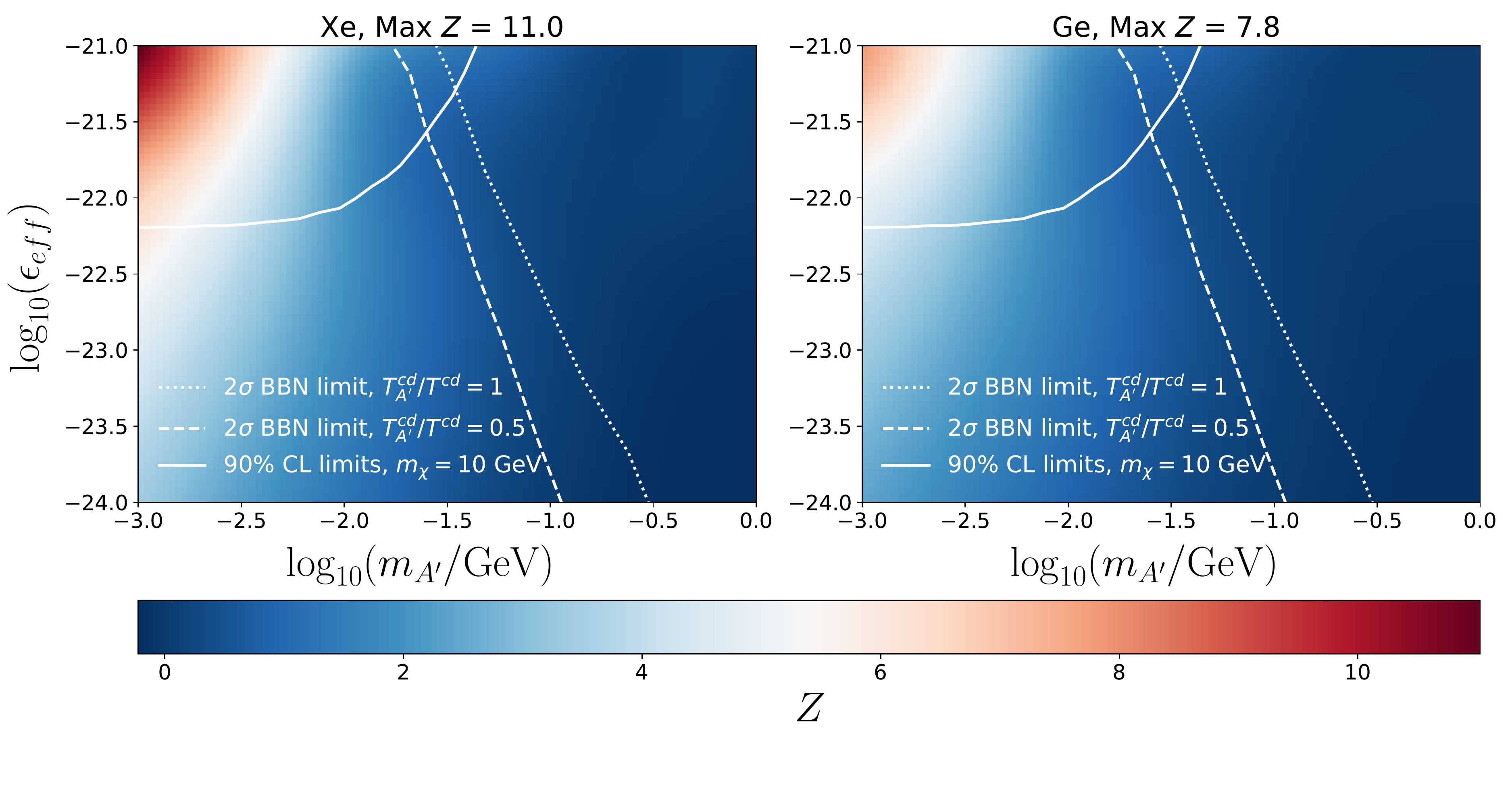}\\
	\includegraphics[scale=0.32]{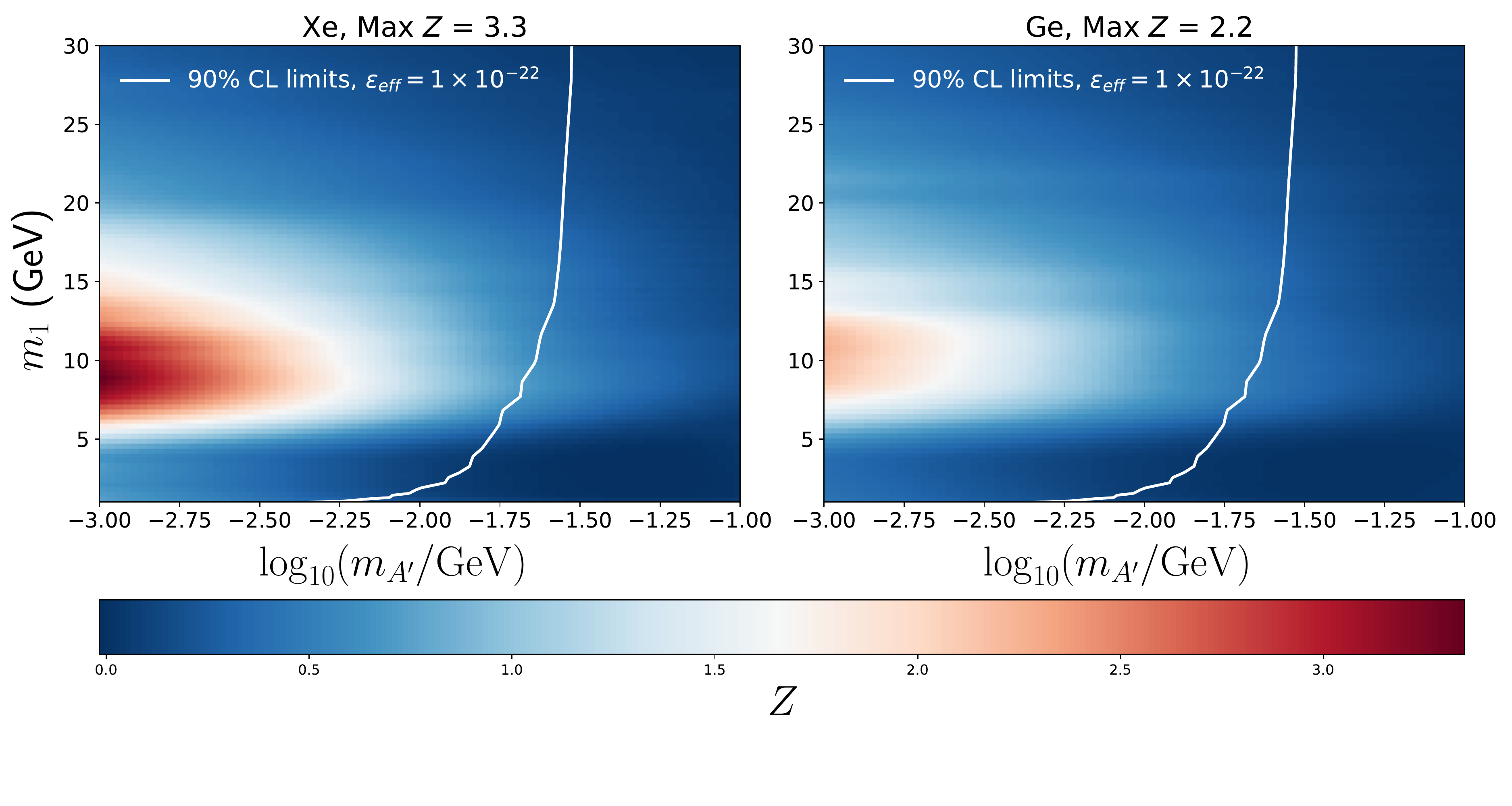}
 	\caption{Median significance $Z$ with which and ADM model with a light mediator can reject the 1DM hypothesis in favour of the 2DM one. \emph{Top:} $m_2-m_1$ plane, with fixed $m_{A^\prime}= 5$ MeV and $\epsilon_\text{eff} = 1\times10^{-22}$. \emph{Middle:} $\log_{10}(m_{A^\prime}/\text{GeV})-\log_{10}(\epsilon_\text{eff})$ plane, with fixed $m_2= 50$ GeV and $m_1 = 10$ GeV. \emph{Bottom:} $\log_{10}(m_{A^\prime}/\text{GeV})-m_1$ plane, with fixed $m_2= 50$ GeV and $\epsilon_\text{eff} = 1\times10^{-22}$. The left panel is for Xe and the right one for Ge. In the last two panels the solid white line is the 90\% CL upper limit on $\epsilon_\text{eff}$ from the latest PandaX results~\cite{Ren:2018gyx}, and in the middle panel the dashed and dotted-dashed lines are the $2\sigma$ \textit{lower} bounds placed on the lifetime of the mediator from BBN~\cite{Hufnagel:2018bjp} for two benchmark $T^\text{cd}_{A^\prime}/T^\text{cd}$ ratios, 0.5 and 1.}\label{fig::light_med_m1m2}
 \end{figure}

In Fig.~\ref{fig::light_med_m1m2} we show the results of the hypothesis testing in the planes $m_2-m_1$ (top), $\log_{10}(m_{A^\prime})-\log_{10}(\epsilon_\text{eff})$ (middle) and $\log_{10}(m_{A^\prime})-m_1$ (bottom). The left panel is for Xe and the right one for Ge. We use 90\% C.L constraints on the $U(1)_\text{DM}$ coupling derived from PandaX~\cite{Ren:2018gyx} to put a conservative upper limit on the parameter combination $\epsilon_\textrm{eff}\equiv\epsilon^2\alpha_\text{DM}\lsim10^{-22}$. These limits are shown as solid white lines in the middle and bottom panels. In the middle panel only the most conservative limit is shown for a DM mass of 10 GeV. Furthermore, a lower limit can be placed on the combination $\epsilon_\text{eff}\,m_{A^\prime}$ from requiring that thermally produced $A^\prime$ particles decay before Big Bang Nucleosynthesis (BBN), with a lifetime below a few seconds~\cite{Lin:2011gj,Kaplinghat:2013yxa,Hufnagel:2018bjp}. We use the $2\sigma$ C.L limits from Ref.~\cite{Hufnagel:2018bjp} which are calculated at two benchmark values for the mediator/decoupling temperature ratio $T^\text{cd}_{A^\prime}/T^\text{cd}$.  These lower limits are overlaid in the middle panels for $T^\text{cd}_{A^\prime}/T^\text{cd}=1$  (dotted white)  and $T^\text{cd}_{A^\prime}/T^\text{cd}=0.5$  (dashed white). We see from these results that, as expected, one needs go to larger $\epsilon_{\text{eff}}$ and smaller $m_{A^\prime}$ to increase the median sensitivity. However the PandaX limits severely restrict the forecasting potential, in the most conservative case allowing a median significance of $Z\sim 2\,(3) $ for Ge (Xe) for $m_{A^\prime}\sim 10$ MeV. That is,  in all cases the regions of highest $Z$ are disfavoured by PandaX null results.

\begin{figure}[t] 
	\centering
	\includegraphics[scale=0.42]{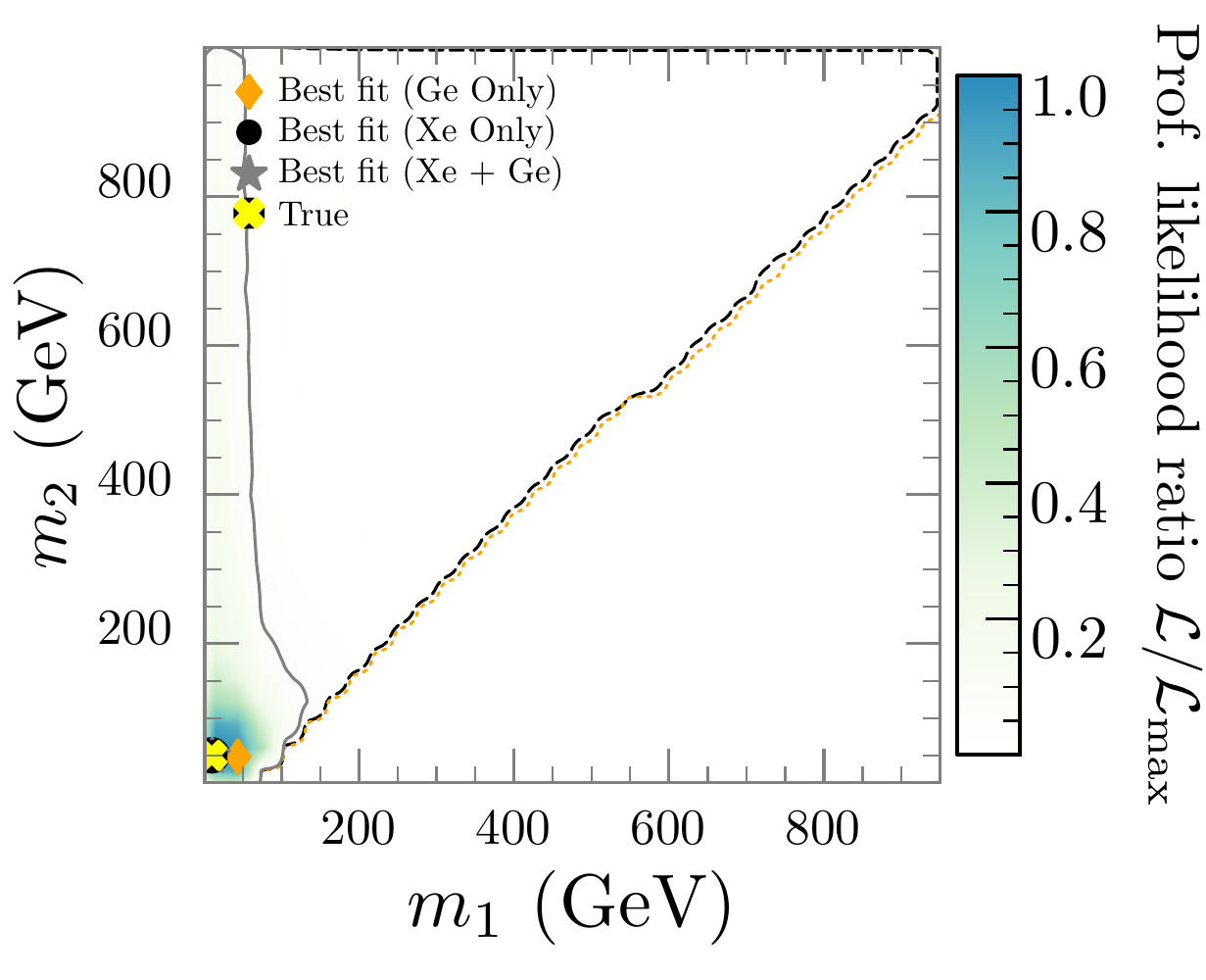}~
	\includegraphics[scale=0.42]{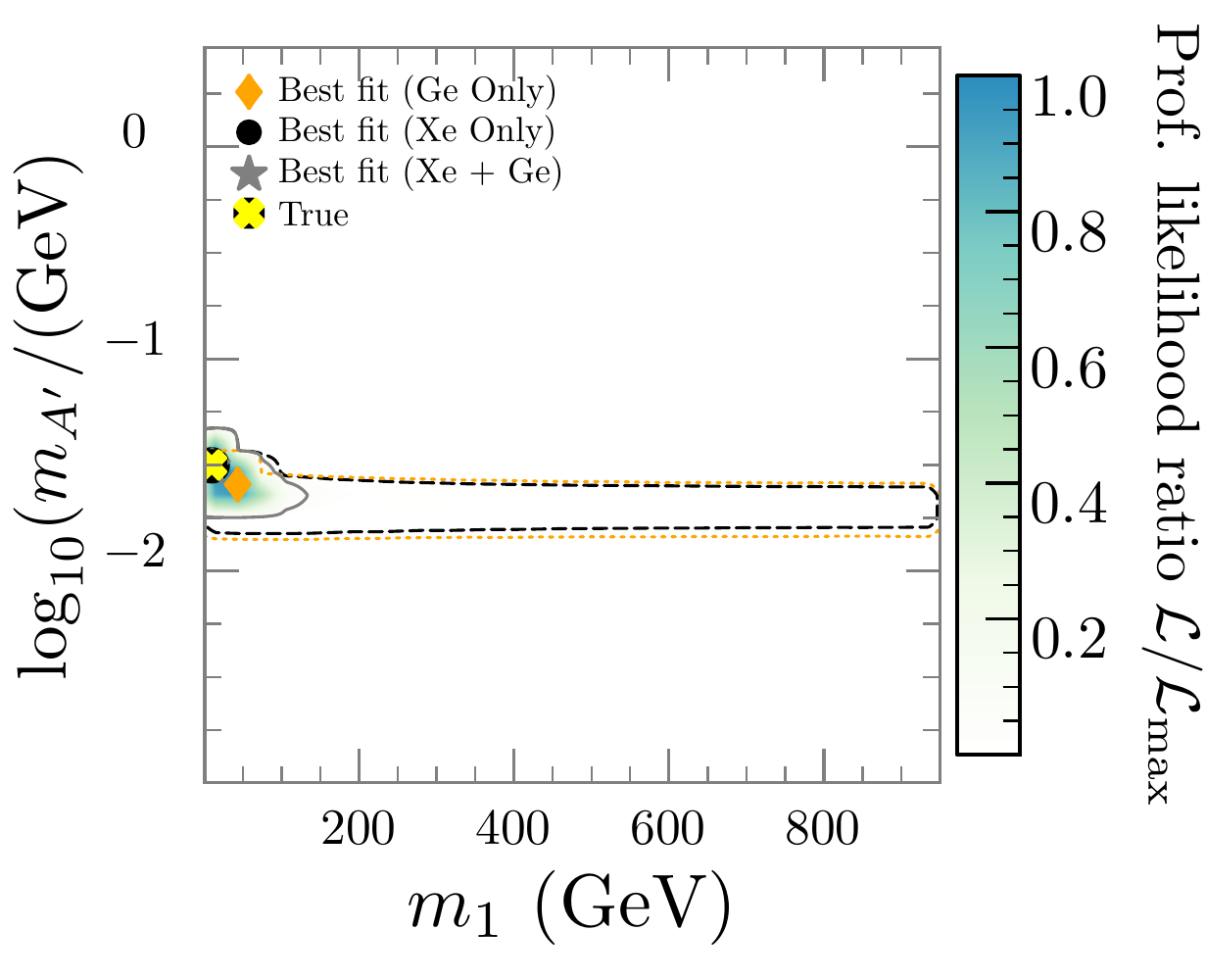}~
	\includegraphics[scale=0.42]{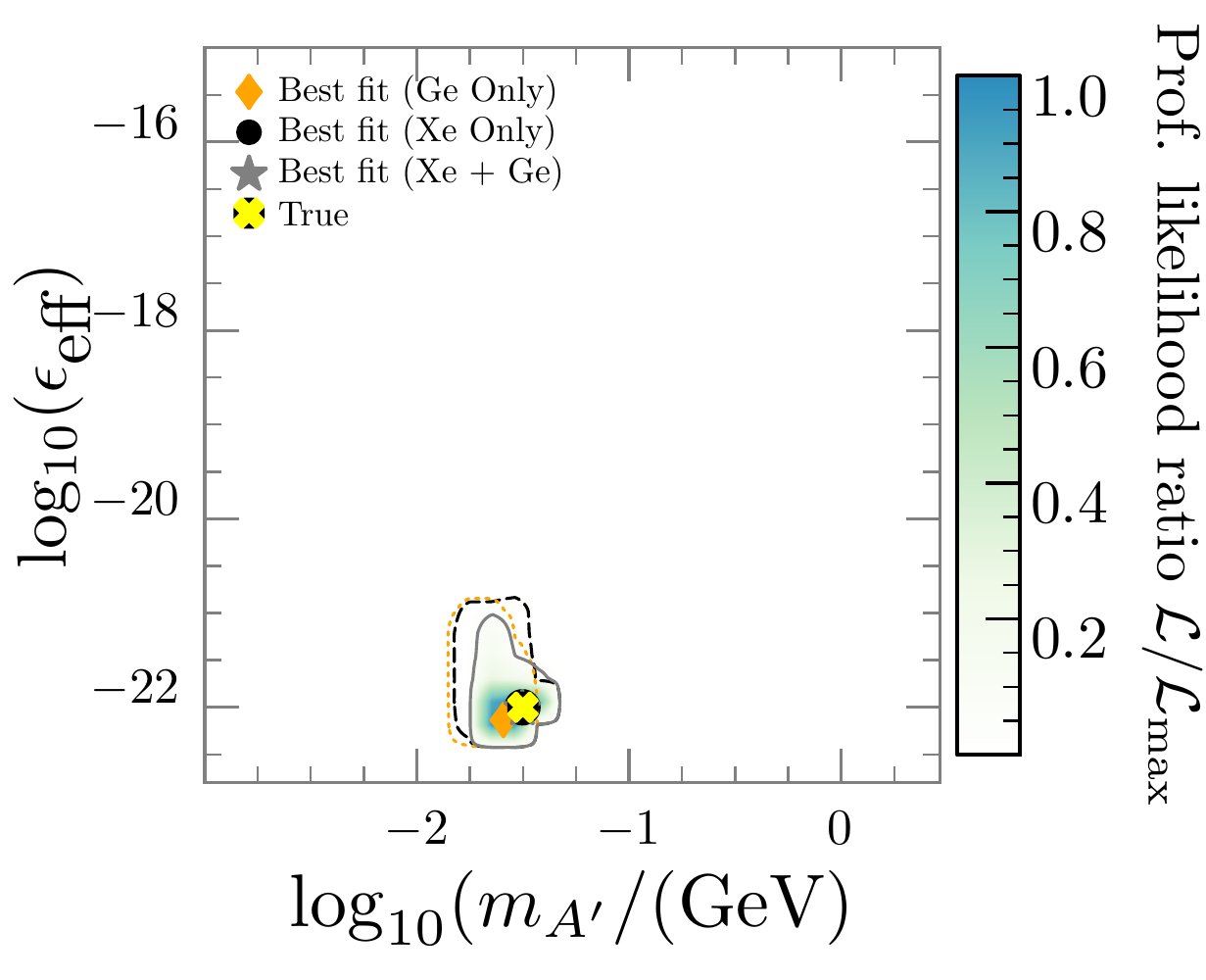}
	\caption{Similar to Fig.~\ref{fig:eq_num_param_est} but for parameter estimation in the ADM scenario with  a light mediator. \emph{Left:} $m_1-m_2$ plane. \emph{Middle:} $m_1-\log_{10}(m_{A^\prime}/\text{GeV})$ plane. \emph{Right:} $\log_{10}(m_{A^\prime}/\text{GeV})-\log_{10}(\epsilon_{\rm eff})$ plane. }
	\label{fig:light_med_param_est}
\end{figure}

In Fig.~\ref{fig:light_med_param_est} we show the results of the parameter estimation for the second benchmark point in Tab.~\ref{tab:benchmarks} in the planes $m_1-m_2$ (left),  $m_1-\log_{10}(m_{A^\prime}/\text{GeV})$ (middle) and $\log_{10}(m_{A^\prime}/\text{GeV})-\log_{10}(\epsilon_{\rm eff})$ (right). We notice a large degeneracy in the $m_1-m_2$ plane. The extended region in the mass of the lightest DM mass, $m_1$, is reduced in the combined Xe+Ge case, however a very large uncertainty remains in $m_2$ at the 2$\sigma$ level.

\section{Thermal freeze-out} \label{sec:thermal_fo}

\subsection{Generic scenarios}  \label{subsec:gen_disc}
Another very natural way to obtain the correct dark matter relic abundance is via thermal freeze-out. In this case, ignoring logarithmic corrections of the DM masses, the relic abundance implies a relation between the thermally averaged annihilation cross sections at freeze-out, that is,
\begin{align} \label{eq:ass_1}
\Omega_1+\Omega_2 = \Omega_\text{DM}\,\,\longrightarrow\,\,
{\langle \sigma_\text{ann} v\rangle}_1^{-1} +
{\langle \sigma_\text{ann} v\rangle}_2^{-1} =
{\langle \sigma_\text{ann} v\rangle}_\text{th}^{-1}\;.
\end{align}
Therefore, the DM particles have similar energy densities if their annihilation cross sections are similar. We can further assume that the local densities of the DM species scale as the cosmological ones, see Refs.~\cite{Bertone:2010rv,Blennow:2015gta}, which for cold DM is a natural expectation~\cite{Anderhalden:2012qt},
\begin{align} \label{ass_2}
\frac{\rho_\beta}{\rho_\text{loc}} = \frac{\Omega_\beta h^2}{\Omega_\text{DM}h^2} 
= \frac{\langle \sigma_\text{ann} v\rangle_\text{th}}{\langle \sigma_\text{ann} v\rangle_\beta}\,,
\end{align}
where we take the canonical value $\langle \sigma_\text{ann} v\rangle_\text{th}=2.2\times10^{-26}$ cm$^3$ s$^{-1}$. $v$ is the DM relative velocity at freeze-out, $v \simeq \sqrt{12/x_f}$, with $x_f \equiv m_{\rm DM}/T_f$. In the following we assume that both DM particles have the same velocity at freeze-out, which is a very good approximation given the fact that $x_f$ only has a logarithmic dependence on the DM mass.

The relationship between the scattering and the annihilation cross sections is model-dependent. We restrict ourselves to scenarios in which the relationship between the two is {simple}. An example is $s$-wave annihilations mediated by some vector mediator into first generation quarks, with SI interactions for scattering. In this case, the annihilation cross section at freeze-out for DM $\beta$ can be written as
\begin{align}
\langle\sigma_\text{ann} v\rangle_\beta  \simeq g_\beta^2\, \frac{m_\beta^2}{(m_\beta^2 + m_{A^\prime}^2)^2}\,v,
\end{align}
where $m_{A^\prime}$ is the mass of the mediator and $g_\beta$ is the effective coupling between quarks and DM particles $\beta$. Note that $s$-wave annihilation is severely constrained by CMB constraints for $m_\beta<\mathcal{O}(10)$ GeV \cite{Ade:2015xua,Berlin:2018bsc} and so we only consider heavy DM in the following. Similarly, the DD WIMP-proton cross-sections have the form
\begin{align}
\sigma_\beta^p (|\vec q|) \simeq g_\beta^2 \frac{\mu_{\beta p}^2}{(|\vec q|^2+m_{A^\prime}^2)^2}\;.
\end{align} 
Hence, neglecting order one factors, we can express the product $r_\sigma\,r_\rho$ in terms of the masses as
\begin{align} \label{eq:rs}
r_\rho\,r_\sigma \simeq \frac{(m_2^2+m_{A^\prime}^2)^2}{(m_1^2+m_{A^\prime}^2)^2\,}\, \frac{\mu_{p2}^2}{\mu_{p1}^2}\,\frac{m_1^2}{m_2^2}\,,
\end{align}
where we have also used Eq.~\eqref{ass_2}. In the following two subsections we  analyse the cases of heavy and light mediators, each of which yields a different phenomenology. 

\begin{figure}[!t]
	\centering
	\includegraphics[scale=0.26]{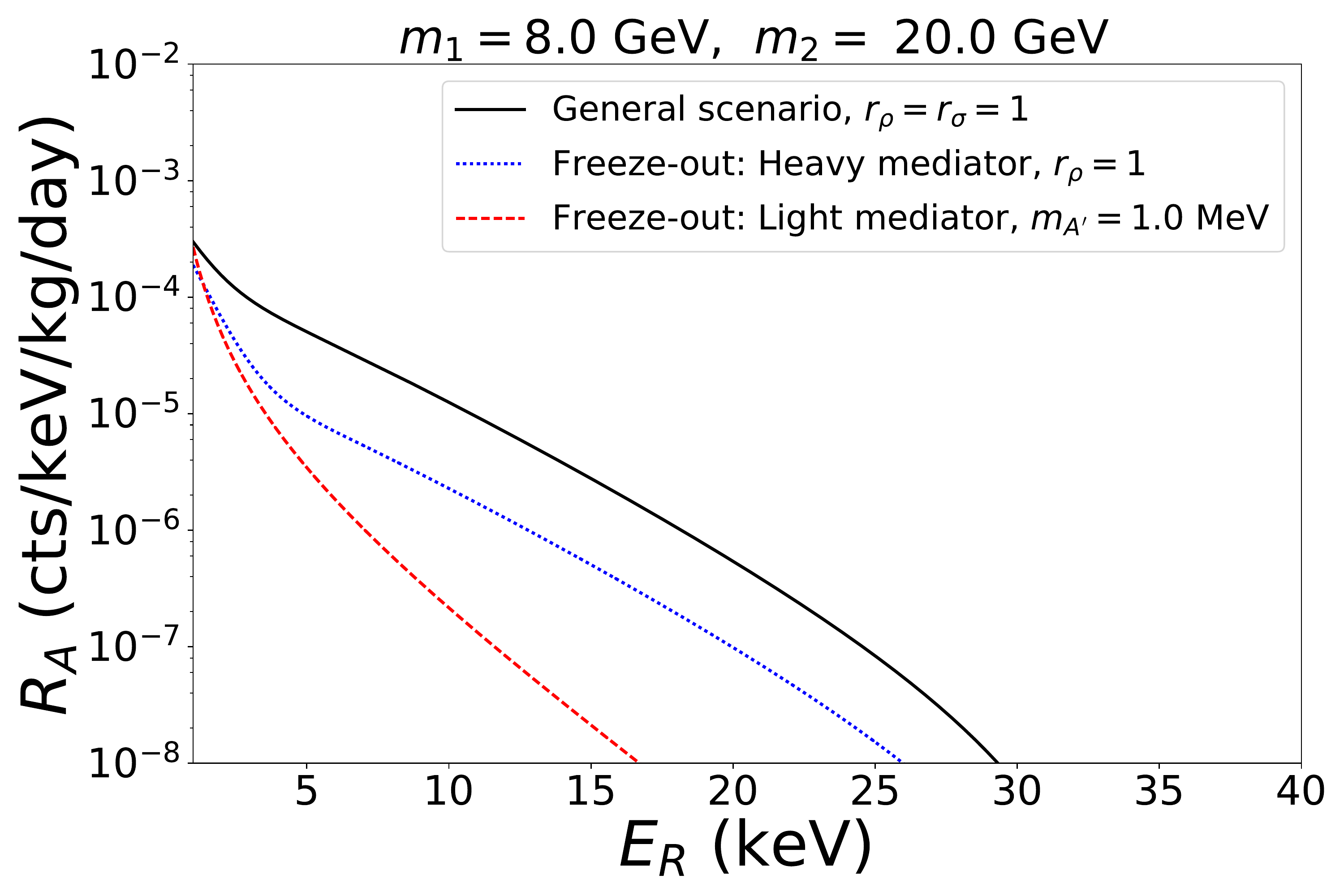}~~
	\includegraphics[scale=0.26]{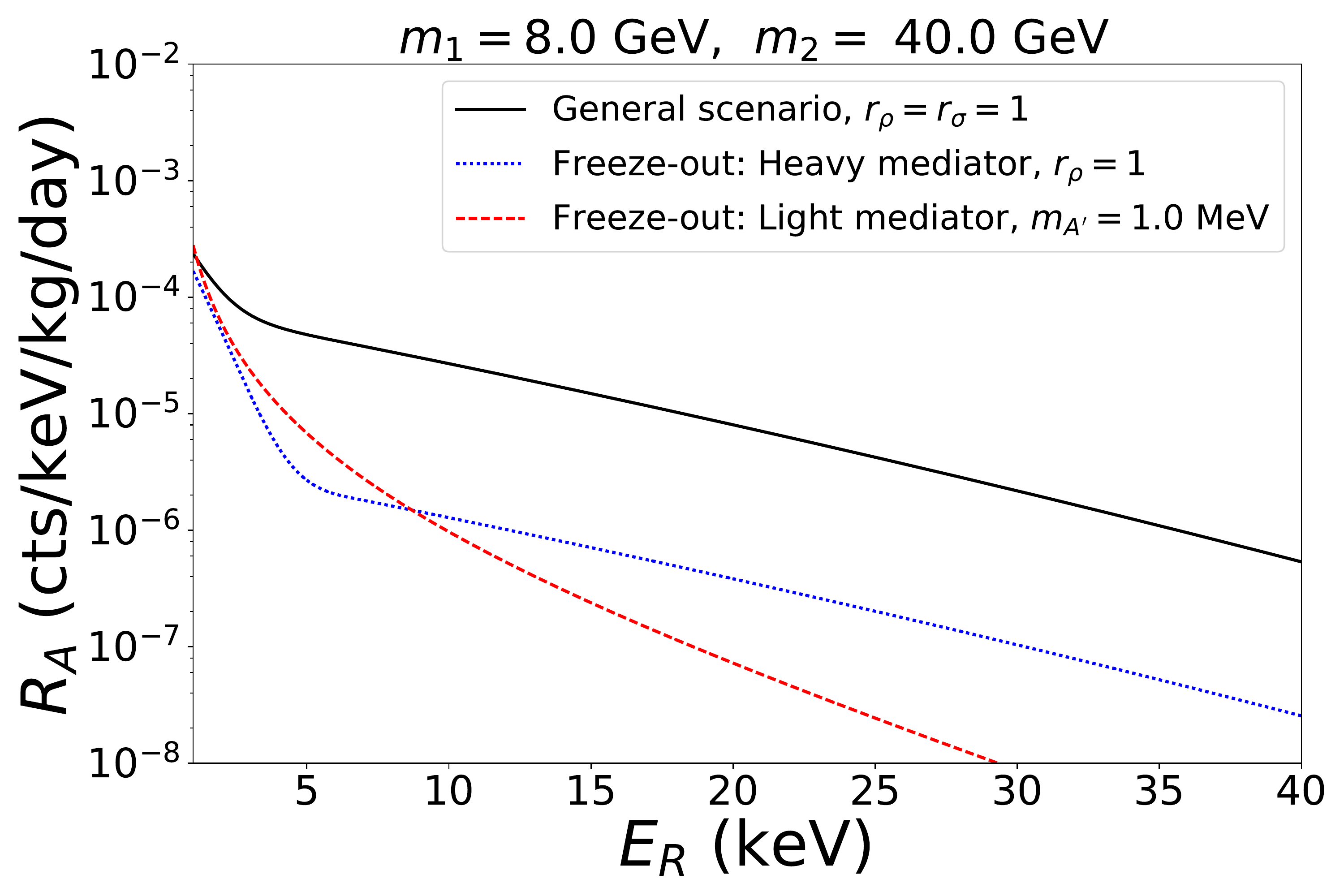}
	\caption{Rates  in a xenon experiment after imposing constraints from thermal freeze-out discussed in this section for the \emph{general} model (solid black), with a heavy mediator (dotted blue) and a light mediator with mass $m_{A^\prime} = 1$ MeV (dashed red). We use  $m_1=8$ GeV. \emph{Left:} $m_2=20$ GeV. \emph{Right:} $m_2=40$ GeV. }
	\label{fig::loc_glob_comp}
\end{figure}

\subsection{Heavy mediators}  \label{subsec:heavy_med}

In the heavy mediator limit, with $m_{A^\prime} \gg m_{1,2}$, we can simplify Eq.~\eqref{eq:rs} to
\begin{align}  \label{eq:rsheavy}
r_\rho\,r_\sigma\simeq \frac{\mu_{p2}^2}{\mu_{p1}^2}\frac{m_1^2}{m_2^2}\;,
\end{align}
and the rate in Eq.~\eqref{eq:rate_tot} reduces to
\begin{align}
\label{eq:heavy_med_rate}
R_{A}(E_{R}) = \frac{\rho_{\rm loc}\,\sigma_1^p}{2 \,(1+r_\rho)\,\mu_{1p}^2\,m_1}\,A^2 \,F_{A}^2(E_{R}) \,\left[\eta(v_{m,A}^{(1)})+\frac{m_1^3}{m_2^3}\,\eta(v_{m,A}^{(2)})\right]\,.
\end{align}
Notice how in this case the contribution of DM2 is suppressed by $m_1^3/m_2^3$. One should therefore expect that the total rate is rapidly dominated by DM1 for $m_2 \gg m_1$. Hence, the discriminating power between 1DM and 2DM in this region of the parameter space should correspondingly be suppressed. Furthermore, the rate is enhanced for small $r_\rho \simeq g_1^2\,m_1^2/(g_2^2\,m_2^2)$ and hence we would expect the median significance to increase. To see this behaviour, we plot in Fig.~\ref{fig::loc_glob_comp} the rate in Eq.~\eqref{eq:heavy_med_rate} as a blue dotted line for two different mass splittings. One should immediately notice the \emph{kink} feature becoming more prominent for $m_1=8$ GeV and $m_2=40$ GeV (right plot).

\begin{figure}[ht]
	
	\centering
	\includegraphics[scale=0.35]{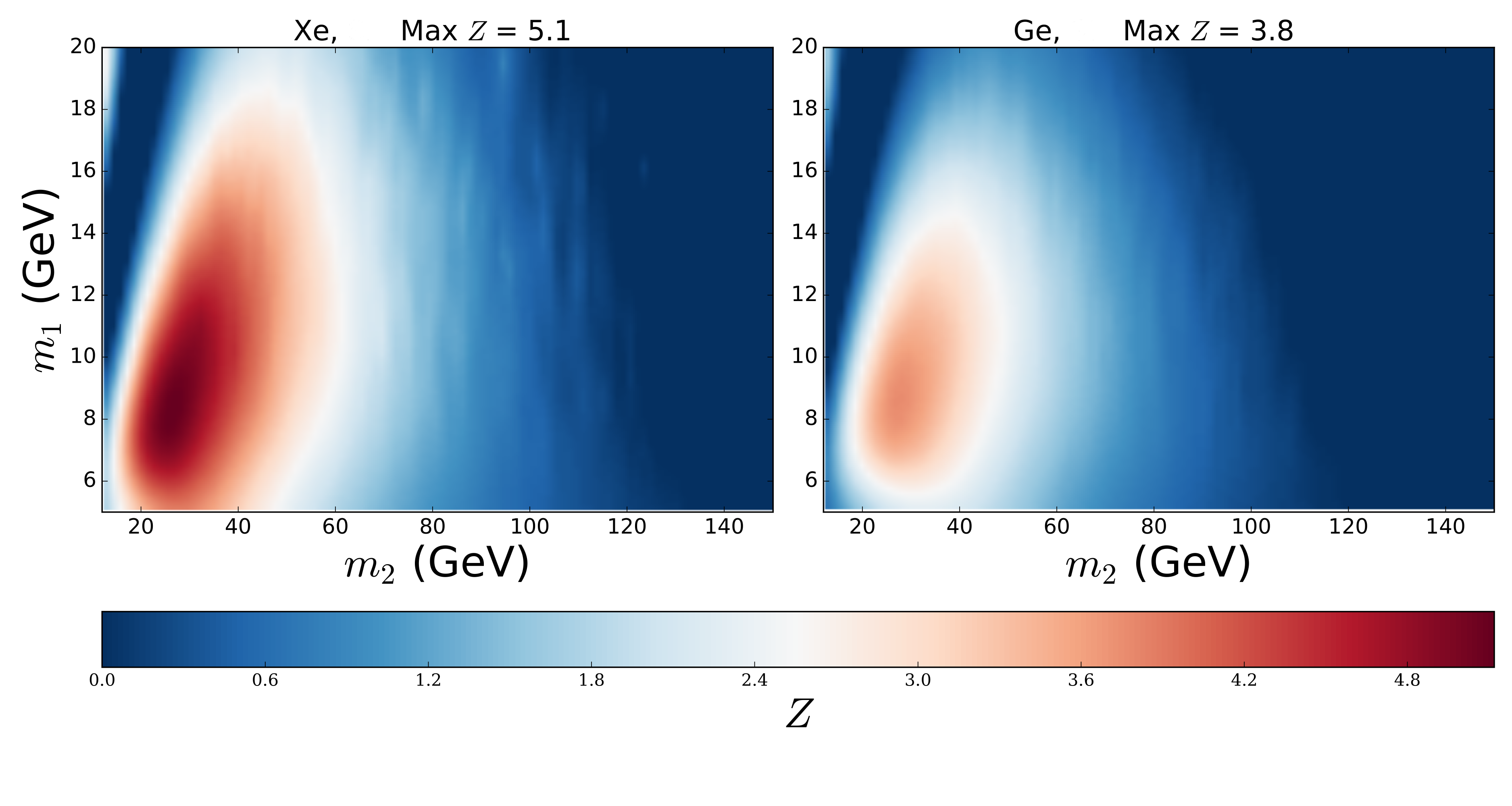}\\
	\includegraphics[scale=0.35]{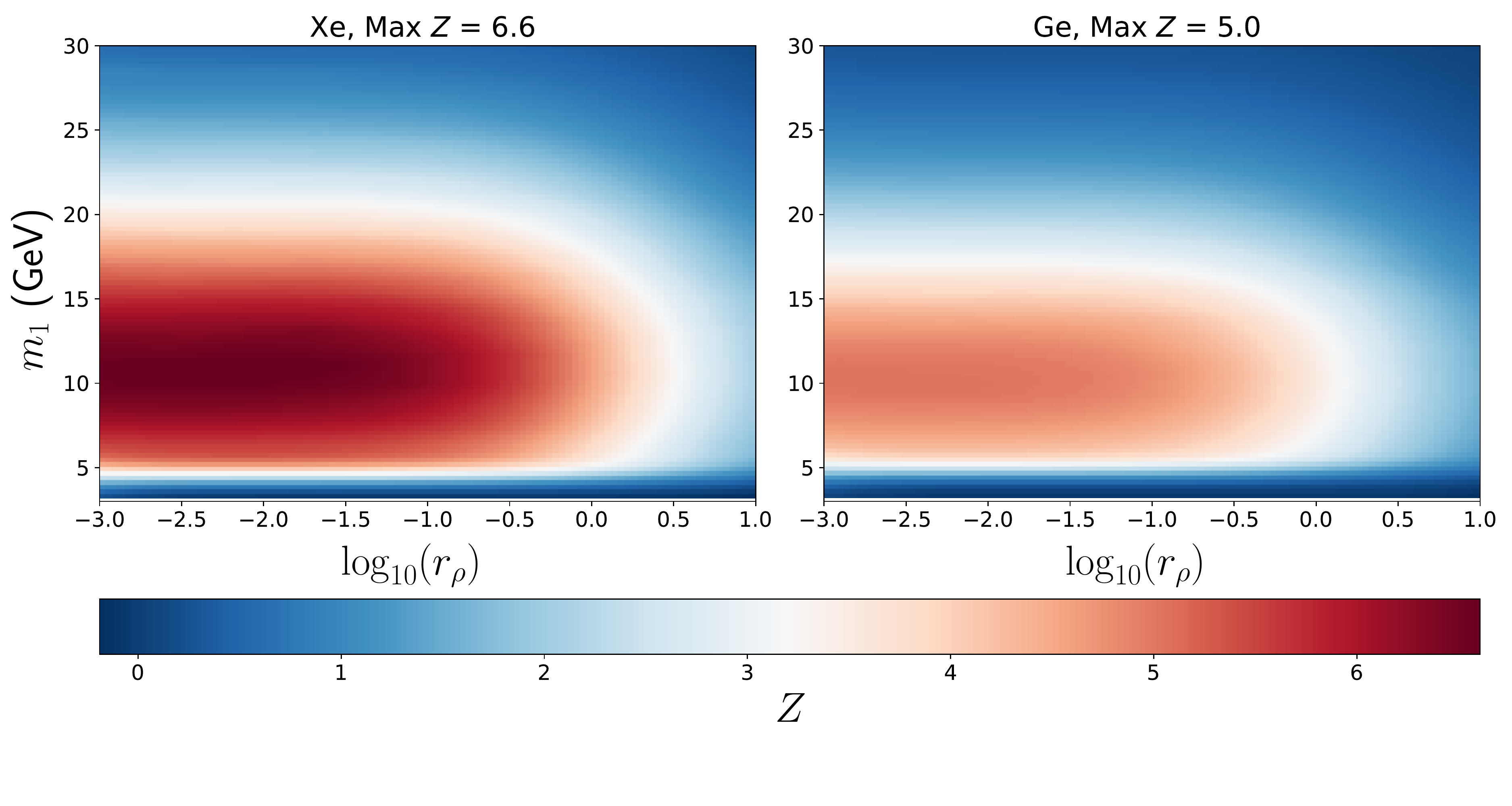}
	\caption{Significance $Z$ for thermal freeze-out with heavy mediators for scenarios where the local density scales as the global one. \emph{Top:} $m_2-m_1$ plane, with fixed $r_\rho= 1$. \emph{Bottom:} $\log_{10}(r_\rho)-m_1$ plane, with fixed $m_2=30$. The left panels are for Xe, and the right ones for Ge.}\label{fig::eq_loc_glob_m1m2} 
\end{figure}

In Fig.~\ref{fig::eq_loc_glob_m1m2} we show the results of the hypothesis testing in the $m_2-m_1$ (top) and $\log_{10}(r_\rho)-m_1$ (bottom) planes for the case of thermally produced DM species coupled to a heavy mediator.  As expected we find that the median significance falls off for large $m_2$ due to the $m_1^3/m_2^3$ suppression factor.  We also observe  in the lower panel an increase and approximate degeneracy in the median significance for vanishingly small $r_\rho$ (i.e., $g_1^2\,m_1^2\ll g_2^2\,m_2^2$) and $m_1\in[5,20]$. The median significance drops for masses below $m_1\sim5$ due to the light component producing scatterings below threshold.   

\begin{figure}[t]
	\centering
	\includegraphics[scale=0.44]{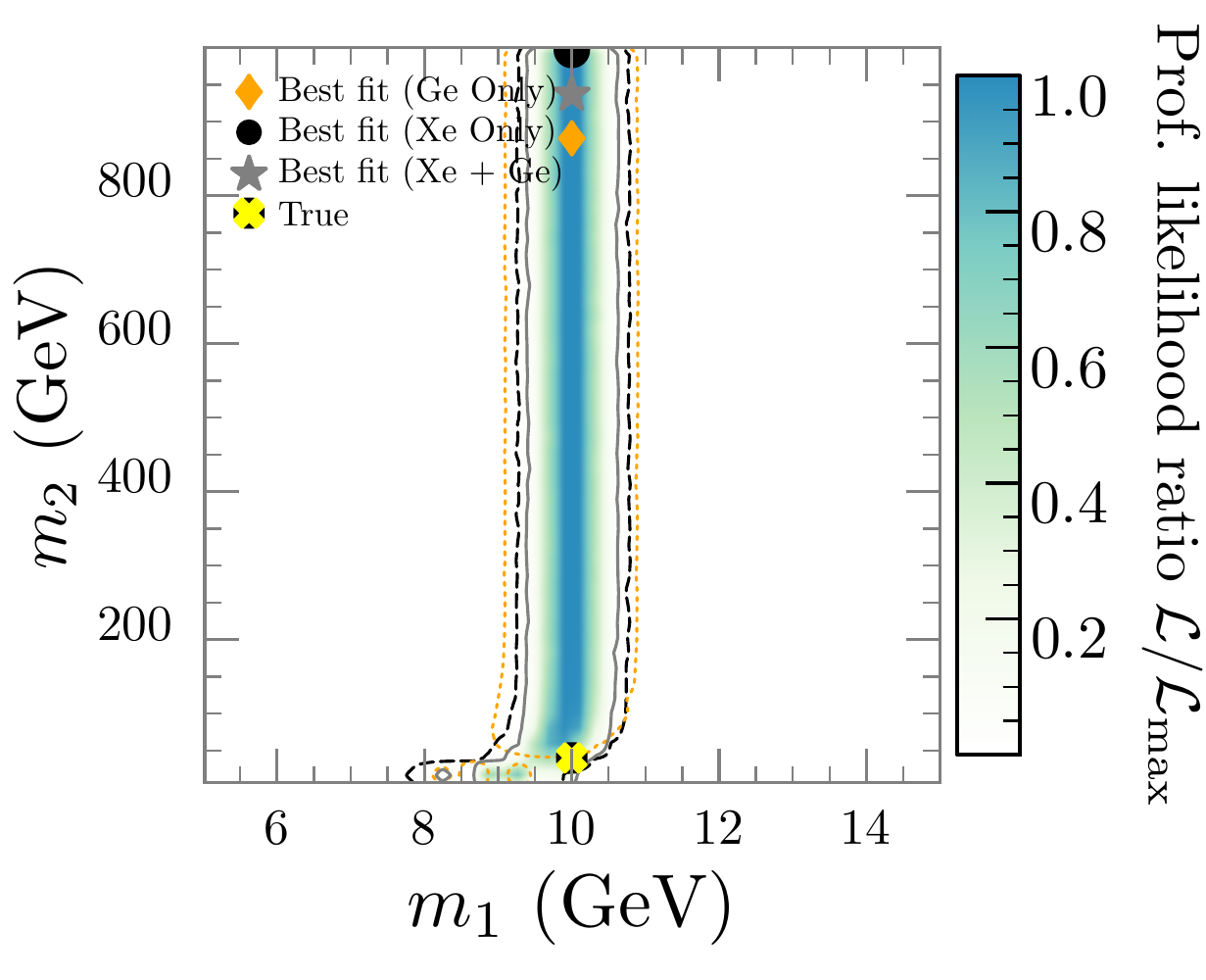}~
	\includegraphics[scale=0.44]{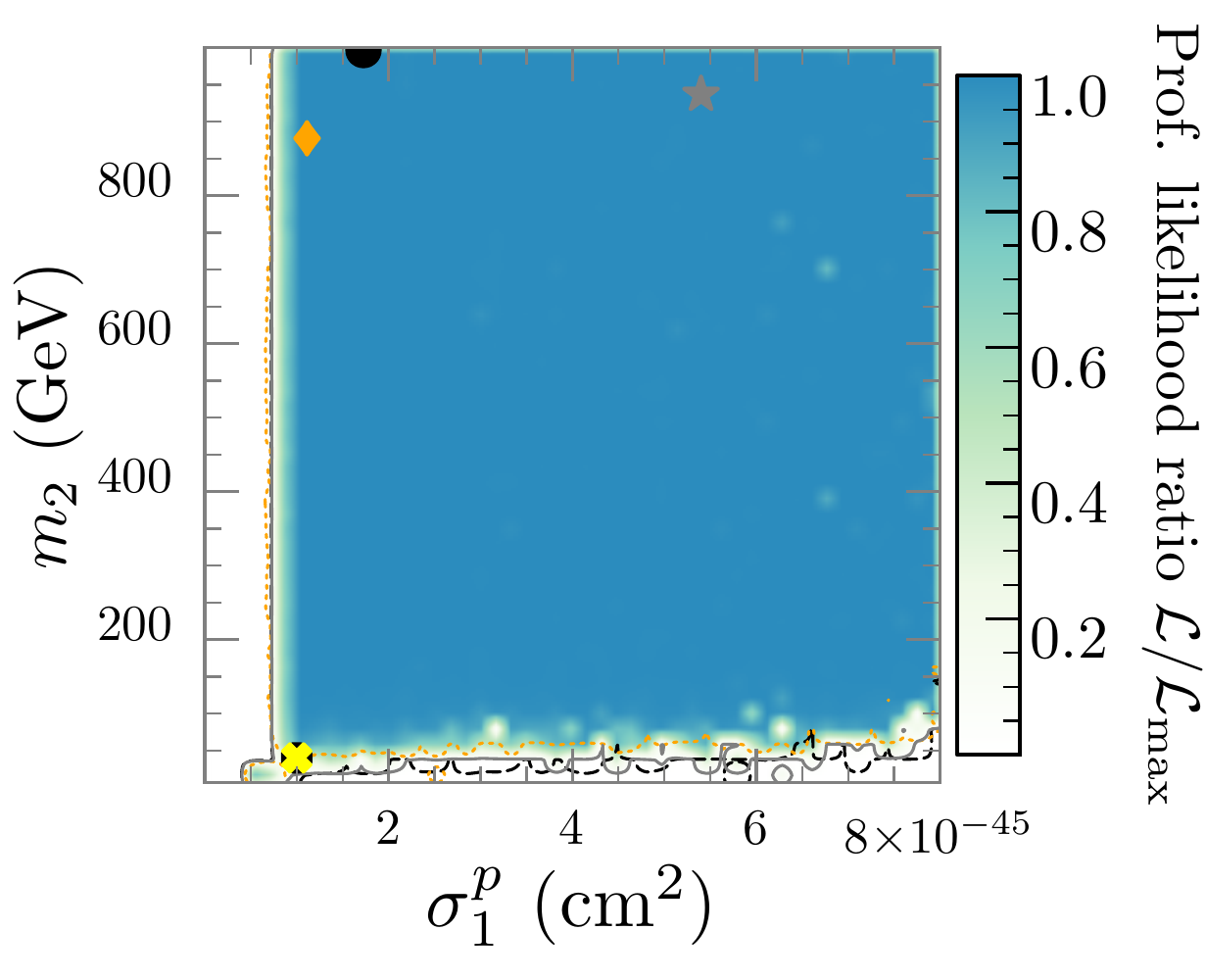}~
	\includegraphics[scale=0.44]{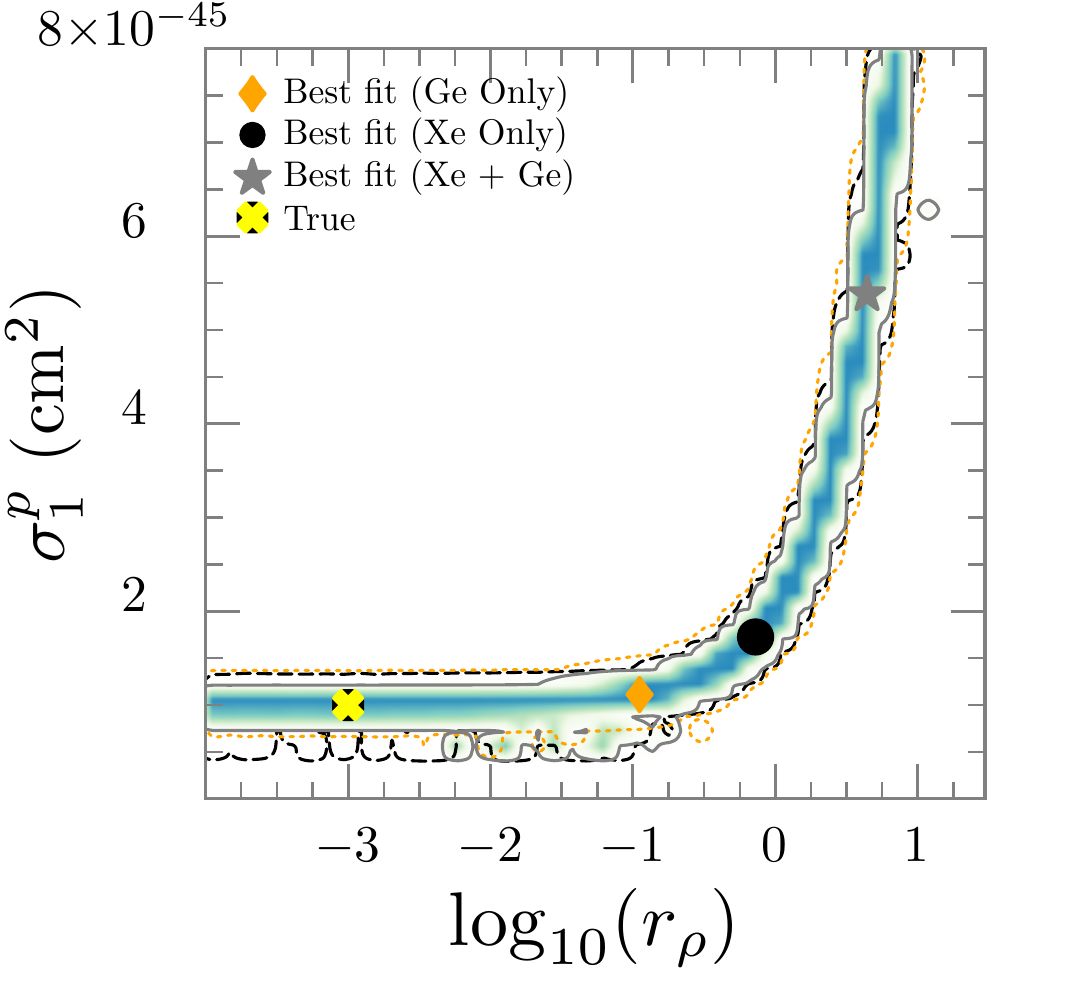}
	\caption{Similar to Fig.~\ref{fig:eq_num_param_est} for parameter estimation for thermal freeze-out models with heavy mediators. \emph{Left:} $m_1-m_2$ plane. \emph{Middle:} $\sigma_1^p -m_2$ plane. \emph{Right:} $\log_{10}(r_\rho)-\sigma_1^p$ plane.}
	\label{fig:loc_glob_param_est}
\end{figure}

In Fig.~\ref{fig:loc_glob_param_est} we show the results of the parameter estimation for the third benchmark point given in Tab.~\ref{tab:benchmarks} in the $m_1-m_2$, $\sigma_1^p-m_2$ and $\log_{10}(r_\rho)-\sigma_1^p$ planes. We observe interesting degeneracies in the PLR. Starting with the left panel, there appears to be an extended degeneracy in the $m_2$ direction for $m_2 \gsim 40$ GeV whilst reconstruction of the true $m_1$ value is good. This is due to the suppressed number of events for $m_2>40$ GeV rendering the rate indistinguishable from a single component. In the middle panel we see no ability in the average experiment to be able to reconstruct the true $\sigma_1^p$ accurately with the degeneracy extending over the entire plane. This is because larger $\sigma_1^p$ drives the overall normalisation of the rate higher, but it does not make the kink more pronounced. The right panel shows perhaps the most interesting degeneracy. We see that the benchmark is found with good precision in $\sigma_1^p$ for $r_\rho\sim 1$, whereas above this value there is a degeneracy with $\sigma_1^p$. This can be easily understood from Eq.~\eqref{eq:heavy_med_rate}, where for $r_\rho>1$ they enter in the normalisation of the rate as $\sigma_1^p/r_\rho$. 

\subsection{Light mediators}
\label{sec::fo_light_med}

In the following we also consider the simplified model with the vector mediator introduced in Sec.~\ref{subsec:light_med}. Up to order one factors, the t-channel $\chi_\beta+\chi_\beta\rightarrow A^\prime+A^\prime$ (secluded) annihilation cross-section into light mediators reads
\begin{align}
\label{eq::light_med_freeze}
\langle\sigma_\text{ann} v\rangle_\beta \simeq Q_\beta^2 \frac{\pi\alpha_D^2}{m_\beta^2}\,v\,,
\end{align}
where we neglected the phase-space factor, as $m_{A^\prime}\ll m_\beta$. Using Eq.~\eqref{eq::sigma0} we can now relate the DD scattering cross-section to the annihilation one,
\begin{align} \label{eq::light_med_xs}
\sigma_\beta^p \simeq \frac{16\,\epsilon^2\,\mu_{p_\beta}^2\,\alpha_{\text{EM}}\:\langle\sigma_\text{ann} v\rangle_\beta\,m_\beta^2}{\alpha_{\text{D}}\,m_{A^\prime}^4\,v}\;.
\end{align} 
Hence, we obtain
\begin{align}
r_\rho\,r_\sigma \simeq \frac{m_2^2}{m^2_1}\frac{\mu_{p2}^2}{\mu_{p1}^2}\,,
\end{align}
where we assumed that both DM particles have similar velocities at freeze-out.
The rate then becomes 
\begin{align}
\label{eq:light_med_rate_thermal}
R_A(E_{R}) = \frac{\rho_{\rm loc}\,\sigma_1^p}{2 \,(1+r_\rho)\,\mu_{p1}^2\,m_1} \,Z^2\,F_{A}^2(E_{R})\,\left[\eta(v_{m,A}^{(1)}) + \,\frac{m_2}{m_1}\,\eta(v_{m,A}^{(2)})\right]\times\frac{m_{A^\prime}^4}{(2m_AE_R+m_{A^\prime}^2)^2}\,.
\end{align}
Interestingly, notice how the DM2 term is now enhanced by $m_2/m_1$, as opposed to the case of heavy mediators in Eq.~\eqref{eq:heavy_med_rate}. The total rate, measured in events/(kg keV day), can we written as (see App.~\ref{app:loc_glob_simple} for details)
\begin{align} \label{eq:Rlight}
R_A(E_{R}) = \frac{15\,\epsilon^2}{\alpha_\text{D}}
 \,F_{A}^2(E_{R})Z^2\,m_1 \,\left[\,\eta(v_{m,A}^{(1)}) +\frac{m_2}{m_1}\,\eta(v_{m,A}^{(2)})\right]\times\frac{1}{(2m_A E_R+m_{A^\prime}^2)^2}\,,
\end{align}
where all the masses are to be evaluated in GeV. The dark fine structure constant $\alpha_\text{D}$ can be expressed in terms of $m_1$ and $r_\rho$ by making use of Eq.~\eqref{eq:ass_1} and charge conservation from Eq.~\eqref{eq::charge_con} as
\begin{align}
\label{eq::alpha_final}
 \alpha_\text{D} = \frac{5\times10^{-3}}{r_\rho}\,\sqrt{1+r_\rho^2}\,\left(\frac{m_2}{100\,\text{GeV}} \right)\,,
\end{align}
which can be plugged into Eq.~\eqref{eq:Rlight}. The rate in Eq.~\eqref{eq:light_med_rate_thermal} is shown in Fig.~\ref{fig::loc_glob_comp} as a red dashed curve for two mass splittings, for a mediator mass of $1$ MeV. As for the case of  a heavy mediator in Fig.~\ref{fig::compare_old_scan_spectra}, there exists a much steeper decrease with recoil energy than for the heavy mediator cases, due to the $1/(|\vec{q}|^2+m_{A^\prime}^2)$ factor in the rate. This smoothens the \emph{kink} in the spectrum, which is the crucial feature needed to provide a good discrimination between the 1DM and 2DM cases. Hence, one should expect that the average experiments capability to discriminate the one and two component cases will be limited unless there is a significant enhancement in the rate from exposure or cross-section.

\begin{figure}[ht]
	\centering
	\includegraphics[scale=0.3]{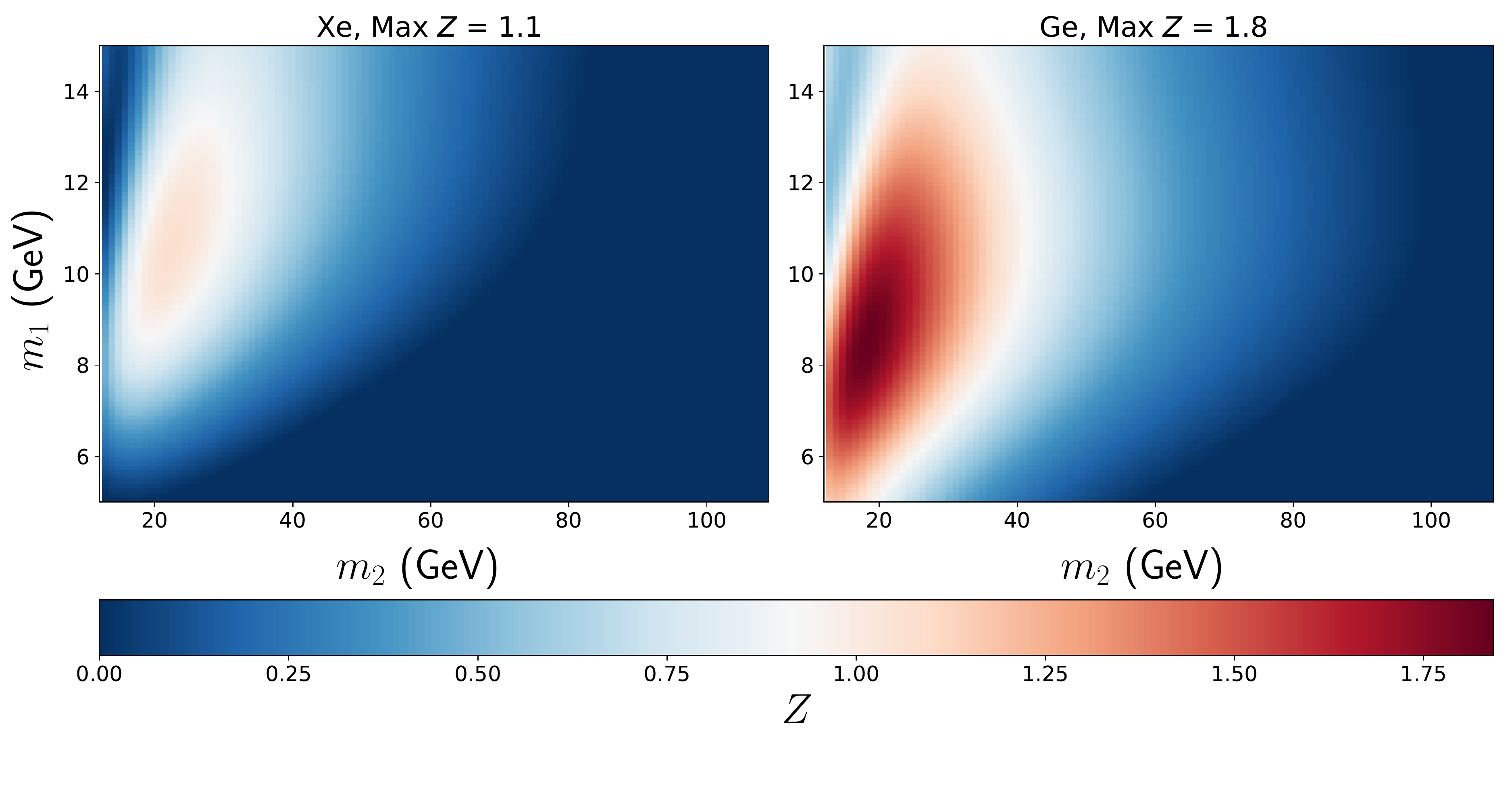}
	\includegraphics[scale=0.3]{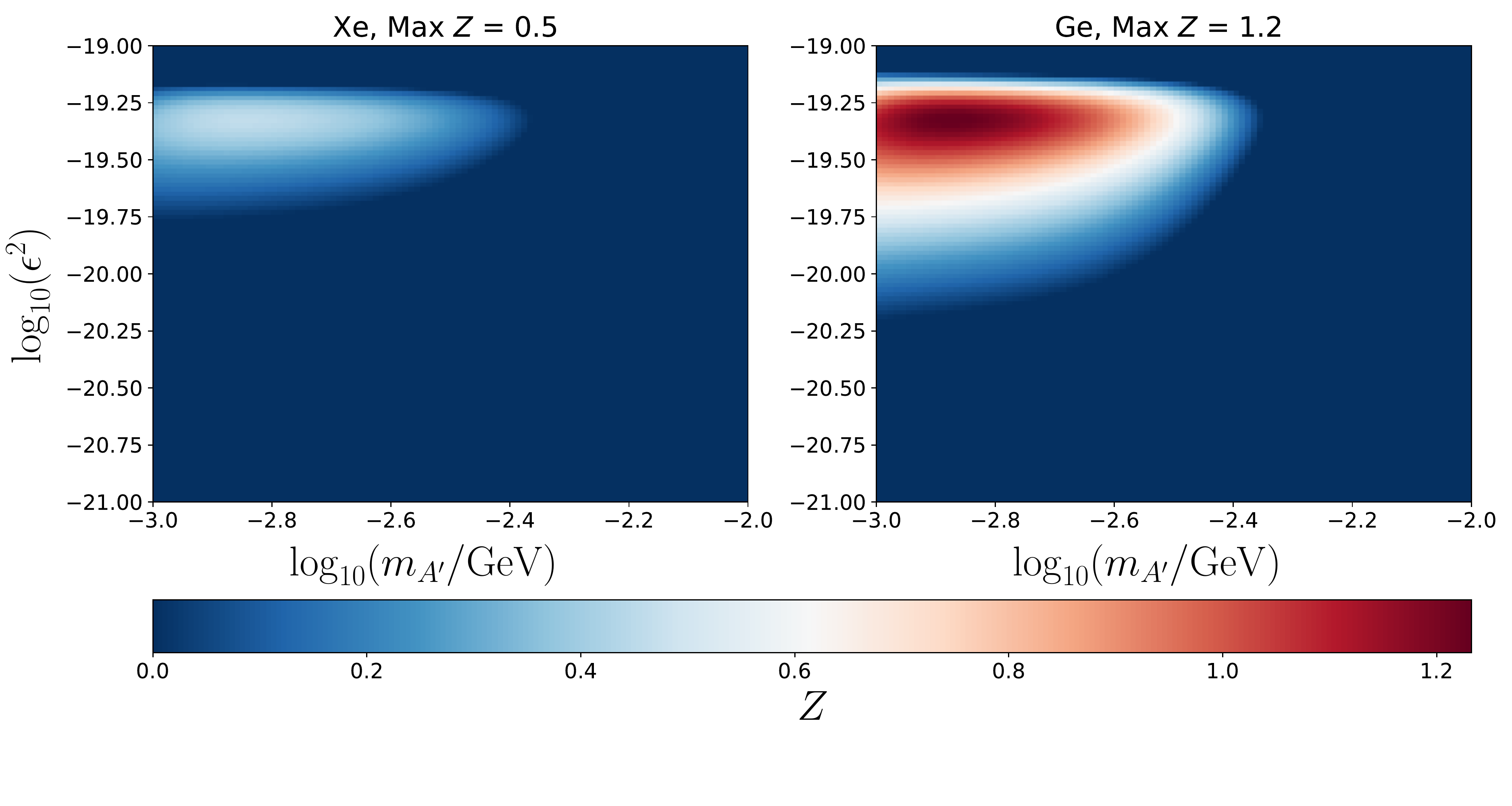}
	\includegraphics[scale=0.3]{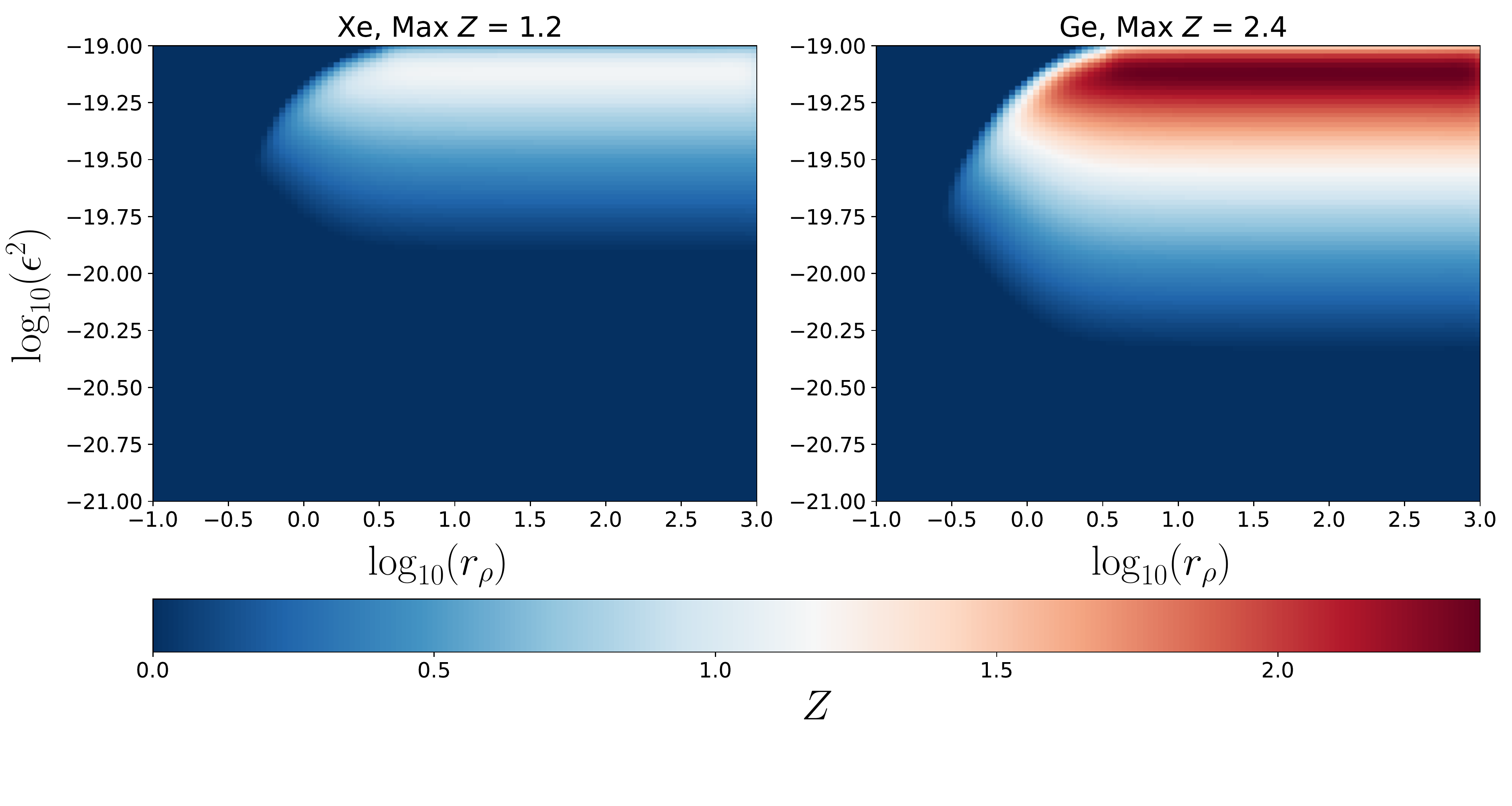}
	\caption{Significance $Z$ of thermal freeze-out models with light mediators. \emph{Top:} $m_2-m_1$ plane with fixed $r_\rho = 1$ and $m_{A^\prime} = 1$ MeV. In the top panel, we dynamically set $\epsilon^2$ to be the maximum allowed by PandaX for each given point in parameter space. The point of maximal $Z$ in the $m_2-m_1$ plane corresponds to $\epsilon^2\sim10^{-21}$.   \emph{Middle:} $\log_{10}(m_{A^\prime}/\text{GeV})-\log_{10}(\epsilon^2)$ plane, with fixed and $r_\rho=1$, $m_2=20$ GeV and $m_1= 8$ GeV. \emph{Bottom:} $\log_{10}(r_\rho)-\log_{10}(\epsilon^2)$, with fixed $m_{A^\prime} = 1$ MeV. Here we set all points that violate the PandaX limit to $Z=0$.  The left panels are for Xe and the right ones are for Ge.}\label{fig::eq_loclightmed_e2_mm}
\end{figure}

For the hypothesis testing shown in Fig.~\ref{fig::eq_loclightmed_e2_mm}, we first perform a scan in the $m_1-m_2$ plane (top) and fix $r_\rho=1$ and $m_{A^\prime} = 1$ MeV. Instead of overlaying the PandaX upper limits as we did earlier in Fig.~\ref{fig::light_med_m1m2} we instead calculate $Z$ at points in the parameter space that are not excluded. That is, for every $m_{2}$ (and $r_\rho=1$) that gives a particular $\alpha_D$ as per Eq.~\eqref{eq::alpha_final},   we set the value of the kinetic mixing parameter $\epsilon^2$ to satisfy $\epsilon^2\,\alpha_\text{D}=10^{-22}$ which is the conservative upper limit. Immediately obvious is the fact that the Ge experiment provides larger median sensitivity than the Xe experiment. Upon inspection of the rates, this feature is due to the fact that at small mediator masses the overall normalisation of the rate goes as $1/(4m_A^2E_R^2)$. We note that the maximum median significance occurs for Ge at $m_1\sim 8$ GeV and $m_2\sim20$ GeV, which corresponds to a maximum allowed $\epsilon^2\sim7\times10^{-20}$.  The median significance completely drops off however above $m_2\sim40$ GeV. This is because the second term in the parentheses in Eq.~\eqref{eq:Rlight} starts to saturate the rate, which becomes 1DM like (i.e., dominated by the heavy component, DM2). Also enhancing this effect is the overall $1/m_2$ suppression in the rate stemming from the $1/\alpha_D$ factor in Eq.~\eqref{eq:Rlight}.

We also show in Fig.~\ref{fig::eq_loclightmed_e2_mm} the planes $\log_{10}(m_{A^\prime}/\text{GeV})-\log_{10}(\epsilon^2)$ (middle) and $\log_{10}(r_\rho)-\log_{10}(\epsilon^2)$ (bottom), for fixed $m_1=8$ and $m_2=20$ GeV. In the general one-component scenario the PandaX limits on the scattering cross-section translate into $\epsilon^2\alpha_D<10^{-22}$. On the other hand, for the model considered here this constraint implies the upper limit $\epsilon ^2/\alpha_D\lesssim 10^{-13} (\text{GeV}/m_1)^2 (1+r_\rho)$ for DM1 and similarly (divided by $r_\rho$) for DM2, where we have corrected for the 2DM local energy densities. All points that do no satisfy this are set to $Z=0$. In the middle panel, we keep $r_\rho=1$ and hence the upper limit on $\epsilon^2$ remains at $7\times10^{-20}$, and hence above this value the median significance is set to zero. We observe in the middle panel that the maximum $Z$ is achieved by making the mediator mass small which is again expected from the $1/(|\vec{q}|^2+m_{A^\prime}^2)$ enhancement in the rate as in the asymmetric case. Furthermore the median significance is maximised for $\epsilon^2\sim5\times10^{-20}$ which is slightly less than the maximum allowed by PandaX. In the bottom panel, we observe that the median significance is maximised and remains constant for $r_\rho > \mathcal{O}(1)$.  This is because unlike other scenarios studied in this paper, the rate in Eq.~\eqref{eq:Rlight} is proportional to the factor $r_\rho/\sqrt{1+r_\rho^2}$, which for $r_\rho<1$ suppresses the rate and therefore $Z$, while for $r_\rho \gg 1$ saturates to $1$. Hence, for large enough $r_\rho$, the rate (and therefore $Z$) falls off to a constant (since the DM masses are fixed). This is indeed what is observed.

\begin{figure}[t]
	\centering
	\includegraphics[scale=0.44]{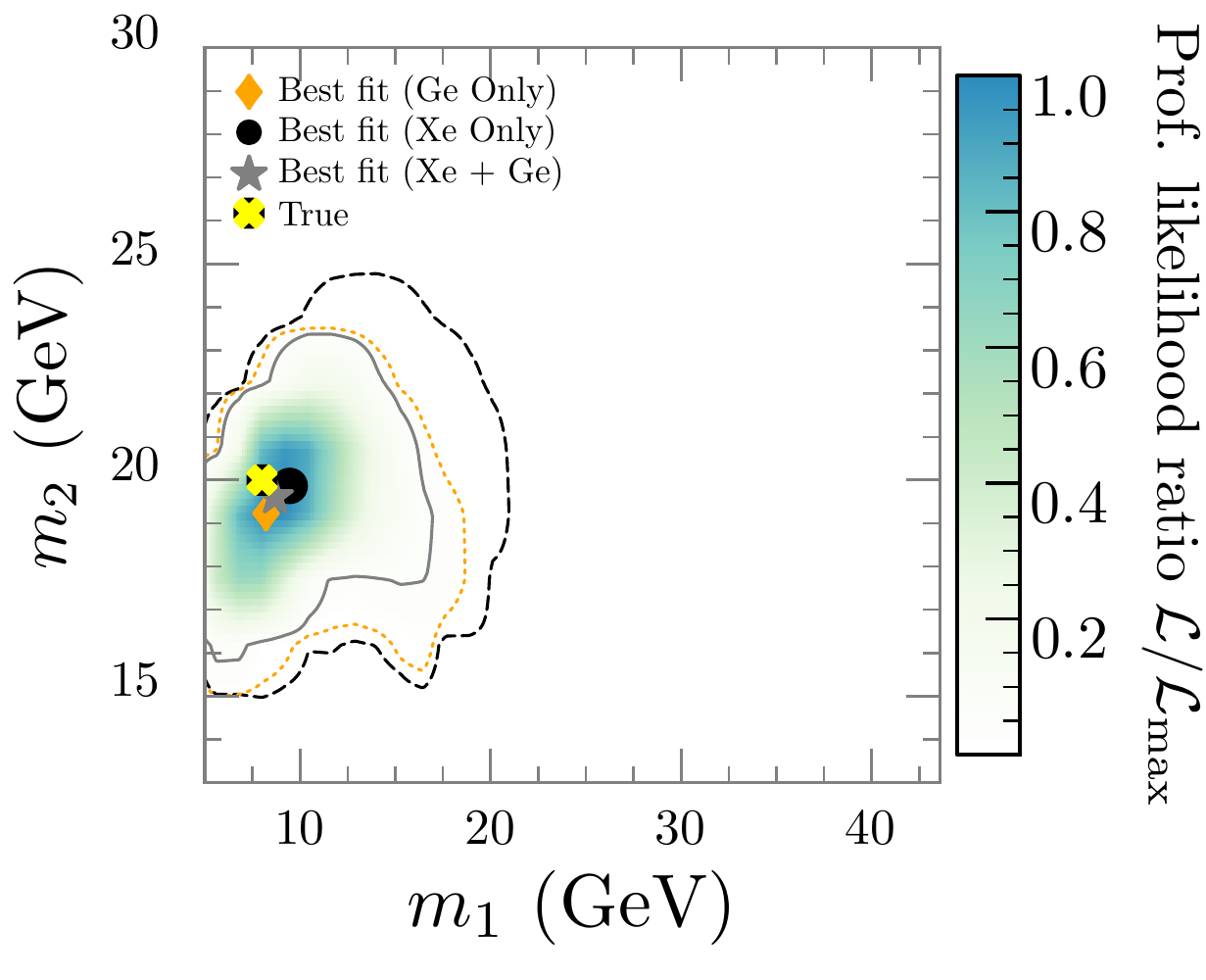}~
	\includegraphics[scale=0.44]{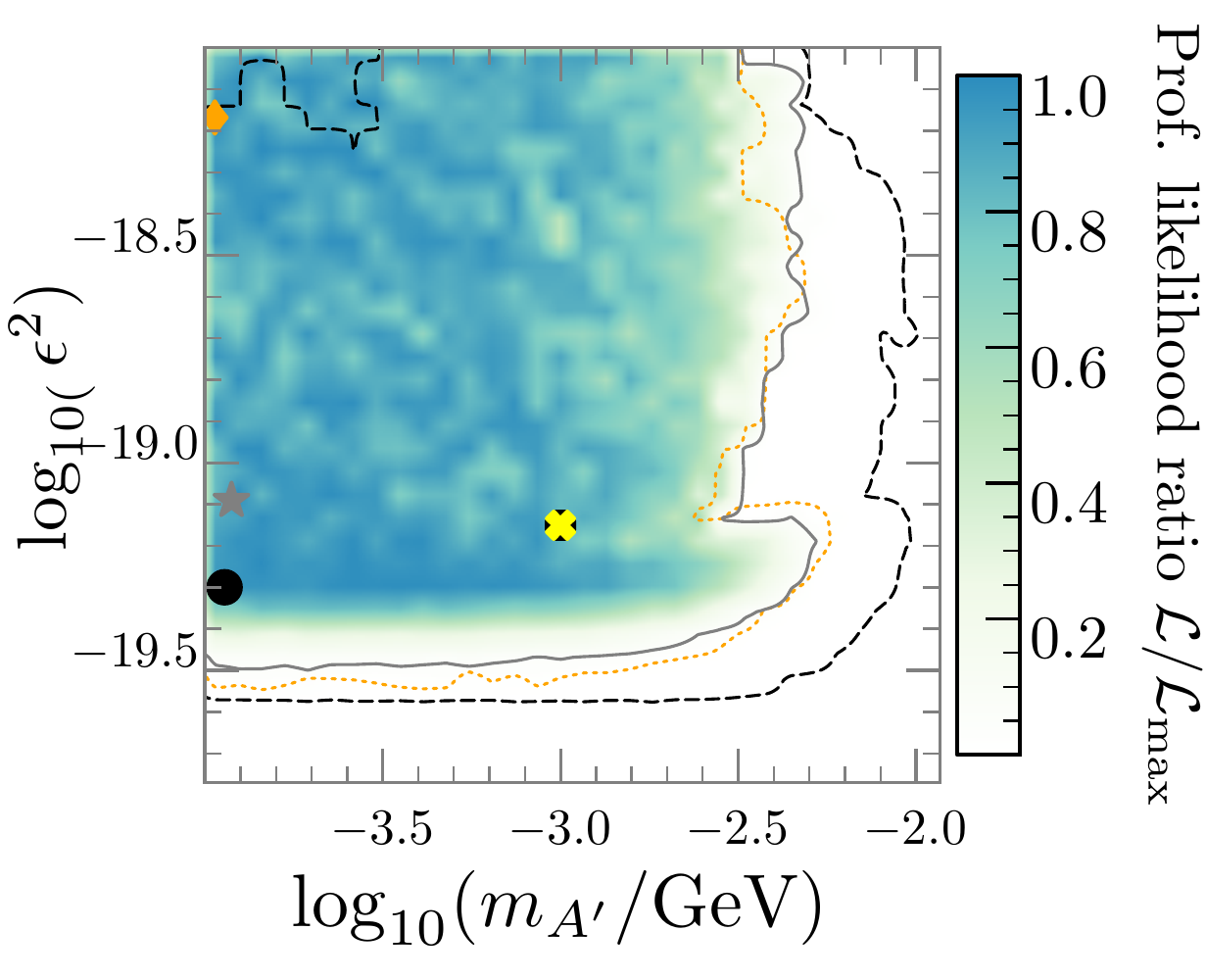}~
	\includegraphics[scale=0.44]{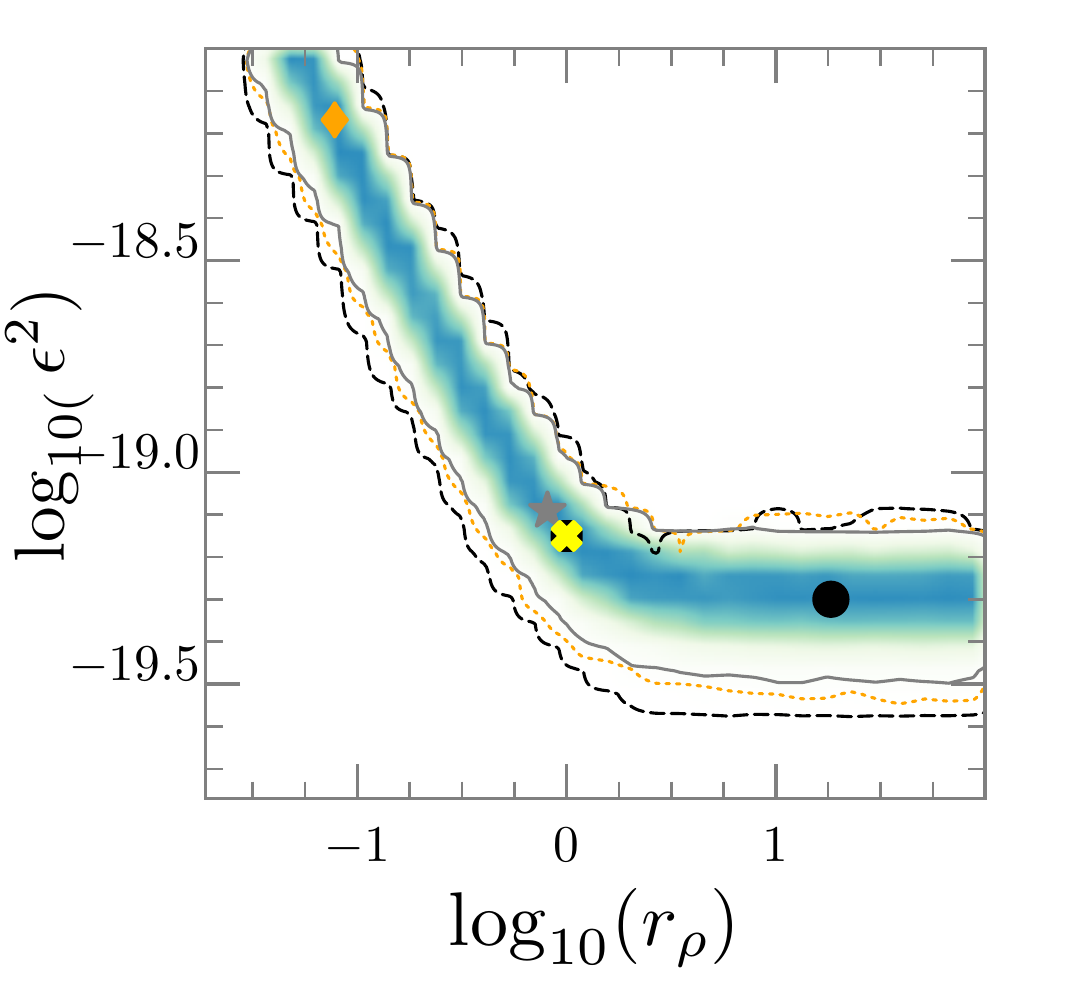}
	\caption{Similar to Fig.~\ref{fig:eq_num_param_est}  for parameter estimation for thermal freeze-out models with light mediators.  \emph{Left: }$m_1-m_2$ plane.  \emph{Middle: } $\log_{10}(\epsilon^2)-\log_{10}(m_{A^\prime})$ plane. \emph{Right: } $\log_{10}(r_\rho)-\log_{10}(\epsilon^2)$ plane.}
	\label{fig:loclight_med_param_est}
\end{figure}

In Fig.~\ref{fig:loclight_med_param_est} we show the results of the parameter estimation for the fourth benchmark point given in Tab.~\ref{tab:benchmarks}. We find that in all displayed cases the germanium experiment provides a tighter 2$\sigma$ C.L region than the xenon experiment. This is because as seen from the hypothesis testing, for this scenario the germanium type experiment provides better maximum median sensitivity. In each panel the combined experiment has a better overall precision as expected. The DM masses are well reconciled with the best fit points finding the true values. The middle plot however shows an extended region of degenerate PLR whilst at the same time the best fit points do not recover the true value.  This is because above $m_{A^\prime}>10^{-2.5}$ GeV the kink becomes less prevalent in the rate, whilst as $m_{A^\prime}\rightarrow 0$ the rate becomes independent of the mediator mass. In the final panel we observe an anti-correlation between $\epsilon^2$ and $r_\rho$ below $r_\rho\sim1$, due to the fact that the rate scales as $\epsilon^2r_\rho$. For $r_\rho>1$ the rate becomes independent of $r_\rho$, hence, as a result there is an extended region where the latter cannot be extracted.

\section{Conclusions} \label{sec:conc}
We have studied the implications that reproducing the relic abundance has on the DD event rates of multi-component DM. We considered two generic genesis scenarios: asymmetric DM and thermal freeze-out. In the asymmetric scenario we restrict the WIMPs to have equal number densities, while in the freeze-out scenario we formulate the analysis based on the assumption that the local energy densities of each component scale like the global relic abundances, so that we are able to relate the thermally averaged annihilation cross-section at freeze-out with the DD WIMP-proton cross-section. For each case we consider two types of mediators, light and heavy. We also analysed the implications on DD of scenarios where the DM components interact with each other, so that their velocity dispersions become mass-dependent. This effect caused a mild smoothing out of the \emph{kink} feature in the two component spectrum. 

For each scenario considered, we first, looked at the shape of the recoil rate spectrum to get an idea of the existence and prominence of any \emph{kink} features which are smoking gun signatures of two-component DM. We then did a hypothesis test to determine regions of the model parameter space where the median experiment can significantly discriminate between the 1DM and 2DM hypothesis. Lastly, we extracted the DM parameters from mock data for a few benchmark model values.

As a general conclusion, we observe a decrease in the maximum median sensitivity relative to the \emph{general} scenario studied in Ref.~\cite{Herrero-Garcia:2017vrl} across all models studied. As a result the maximum median significance to reject the 2DM hypothesis in favour of the 1DM one is suppressed relative to the latter.  

In the asymmetric scenario with heavy mediators, we found that imposing equal number densities smoothens the \emph{kink} feature in the rate relative to the \emph{general} scenario. We also observed that the point that maximises the median sensitivity is at lower $m_2$ than the \emph{general} case. We found that the median experiment is able to sufficiently reconstruct benchmarks in regions of good model discrimination without the presence of any degeneracies. The second model considered in the asymmetric scenario involved the addition of a light vector mediator. We observed that the current PandaX limits on the kinetic mixing parameter and dark $U(1)_\chi$ coupling exclude regions of the model parameter space that give the best hypothesis discrimination. We also observed that the limit on the mediator lifetime from BBN considerations also places a strong constraint on the allowed parameter space.  We obtained extended uncertainties in the resolving power of $m_1$, but the true benchmark had the largest PLR.         

For the case of freeze-out, we observed that the rate for the case of a heavy mediator has a deeper and more pronounced \emph{kink} at low recoil energies relative to the \emph{general} case, whereas the light mediator case still has a smoothened rate.  With the heavy mediator, we find that the median sensitivity is largest for $r_\rho<1$, whilst the $m_1^3/m^3_2$ suppression factor in the rate of the second component causes the median significance to be maximised for smaller $m_2$. In this scenario we find that parameter reconstruction of the heavy DM mass and the cross-section have large associated uncertainties. For the case of the light mediator we considered secluded annihilation to a light vector species that produces a thermally averaged cross-section that is independent of the kinetic mixing parameter $\epsilon$ and is only dependent on the dark coupling $\alpha_D$. Regardless of parameter configuration the spectrum in this case is a smooth curve as opposed to the clear \emph{kink} needed to successfully discriminate between 1DM and 2DM. We find that the maximum median sensitivity is suppressed for $r_\rho < 1$. Our analysis also shows that the DM masses can be recovered whilst the kinetic mixing parameter $\epsilon^2$ experiences some interesting degeneracies with the mediator mass $m_{A^\prime}$ and $r_\rho$. 

To summarise, in this study we found that even when applying reasonable DM genesis model constraints to a general two-component DM scenario, therefore reducing the allowed parameter space, there are cases where the 1DM and the 2DM hypothesis can be significantly discriminated, but typically with smaller significance than for the most general case. We have also uncovered interesting degeneracies in the median experiments capability for reconstructing the model parameters in these scenarios, giving insight into the reconstruction of such models in future direct detection experiments.

\vspace{1cm}

{\bf Acknowledgements\\}
This work is supported by the Australian Research Council through the Centre of Excellence for Particle Physics at the Terascale CE110001004. MW is supported by the Australian Research Council Future Fellowship FT140100244. 

\newpage 
\appendix

\section{Analysis methods} \label{app:analysis}
The analysis methods used in this work are detailed in Ref.~\cite{Herrero-Garcia:2017vrl}. Here for completeness we briefly review the main aspects.

\subsection{Hypothesis testing}
\label{app:hypothesis}
For assessing the average capability of an experiment for discriminating 1DM (modelled by parameters $\theta_{\Ho}$) from 2DM (modelled by parameters $\theta_{\Ht}$) we formulate our analysis as a frequentist hypothesis test with the null-hypothesis being 1DM, and the alternative hypothesis being 2DM. We use a common test statistic constructed from the difference of $\chi^2$:
\begin{equation}
\label{TestStat}
\mathcal{T}= \min_{\theta_{\Ho}} \chi^2(\theta_{\Ho}) -\min_{\theta_{\Ht} }  \chi^2(\theta_{\Ht})\,.
\end{equation}
Notice that the definition of $\mathcal{T}$ is such that the larger its value, the larger the preference for $\Ht$, and the smaller its value, the more $\Ho$ is preferred. The median significance $Z$ quantifies an average experiments capability to discriminate a 1DM hypothesis from a 2DM one. It is defined as
\begin{align}
\label{nsigma}
Z(p) = \sqrt{2}\,\rm erfc^{-1}(p)\;,
\end{align} 
where $p$ is a p-value given by
\begin{align}
\label{p-val}
p = 1-\rm CDF^{\chi^2}_{k\;\rm d.o.f}(\T^{\rm2DM}_0)\;,
\end{align}
and $\T^{\rm2DM}_0$ is the `Asimov likelihood' defined by
\begin{equation}
\label{T0}
\mathcal{T}^{\rm 2DM}_0 \equiv \mathcal{T} (x_i=\mu_i (\theta_{\Ht}^{\rm true}))= \min_{\theta_{\Ho} } \sum^n_i \,\left(\frac{\mu_i (\theta_{\Ht}^{\rm true})-\mu_i (\theta_{\Ho})}{\sqrt{\mu_i (\theta_{\Ht}^{\rm true})}} \right)^2\,.
\end{equation}
The `Asimov data' $\theta^\text{true}$ is generated for Xe and Ge type detectors.

\subsection{Parameter estimation}
\label{app:param_est}
For examples of parameter configurations that provide good discrimination (high $Z$) we conduct parameter estimates. For this we use the method of maximum likelihood which involves calculating the profile likelihood ratio (PLR)
\begin{align}
\label{PLR}
\lambda (\theta_1,\theta_2) = \frac{\L(\mathbf{x}\,|\,\theta_1,\theta_2,\hat{\hat{\theta_3}}...\hat{\hat{\theta_{n}}})}{{\L(\mathbf{x}\,|\,\hat{\boldtheta})}} \equiv\frac{\L(\mathbf{x}\,|\,\theta_1,\theta_2,\hat{\hat{\theta_3}}...\hat{\hat{\theta_{n}}})}{{\L_{\rm max}}} \;.
\end{align}
In our analysis we call the point in the $\theta_1$-$\theta_2$ plane that maximises the profile likelihood ratio the `best-fit' point.  In the limit of large statistics, the distribution of $-2\ln\lambda(\theta_1,\theta_2)$ tends towards a $\chi^2$ with $k=2$ degrees of freedom as given by Wilk's theorem. As a result, this leads to a critical region that is defined by a cut on $\lambda$ that we choose to represent contours of the standard 2$\sigma$ (95.45\%) frequentist confidence level (C.L). For this analysis we use Gaussian likelihoods where we again make use of the Asimov data  $\boldtheta^{\rm true}$:
\begin{align}
\label{AmimovLikePS}
\L(\mathbf{x} = \mu (\boldtheta^{\rm true})\,|\,\boldtheta) =\prod\limits^N_i\, \frac{1}{\sqrt{2\pi\mu_i (\boldtheta^{\rm true})}}e^{-\frac{\left[\mu_i (\boldtheta^{\rm true})-\mu_i (\boldtheta)\right]^2}{2\mu_i (\boldtheta^{\rm true})} }\;.
\end{align}
Visualisation is done with the \texttt{Pippi} plotting package \cite{Scott:2012qh}. We show the best-fit point and normalised profile likelihood density $\L/\L_{\rm max}$ on the colour scale.

\section{Expressions for freeze-out scenarios with light mediators} \label{app:loc_glob_simple}

We can simplify the pre-factor in Eq.~\eqref{eq:light_med_rate_thermal} by using Eq.~\ref{eq::light_med_xs},
\begin{align}
\frac{\rho_{\rm loc}\,\sigma_1^p}{2 \,(1+r_\rho)\,\mu_{1p}^2}
 &= \frac{2.5\times10^{-38}\,m_1^2}{m_{A^\prime}^4}
 \frac{\epsilon^2}{\alpha_\text{D}} \left(\frac{\rho_\text{loc}}{0.4\,\text{GeV cm}^{-3}}\right)
\left(\frac{\langle  \sigma_\text{ann}v\rangle_\text{th}}{2.2\times10^{-26}\,\text{cm}^3 \text{ s}^{-1}}\right)\,\text{GeV}^{-1}\,\text{cm}^{-1} \;,
\end{align}
where we used $v = 0.7c$ and the masses are measured in GeV. Then the total rate in Eq.~\eqref{eq:light_med_rate_thermal}, measured in events/(kg keV day), can easily be written as in Eq.~\eqref{eq:Rlight}. Furthermore, the dark fine structure constant $\alpha_\text{D}$ can be expressed in terms of $m_1$ and $r_\rho$ as in Eq.~\eqref{eq::alpha_final}. This can be checked by making use of Eq.~\eqref{eq:ass_1}, which using Eq.~\eqref{eq::light_med_freeze} reduces to 
\begin{align}
\label{eq::alpha_deriv}
\frac{1}{\alpha_\text{D}^2\,v}\left(\frac{m_1^2}{Q_1^2} +\frac{m_2^2}{Q_2^2} \right)= \frac{1}{\langle \sigma_\text{ann} v\rangle}_\text{th}\quad \longrightarrow \quad\alpha_\text{D} = 5\times10^{-5}\, \sqrt{\frac{m_1^2}{Q_1^2} +\frac{m_2^2}{Q_2^2} }\, \text{GeV}^{-1}\,,
\end{align}
with the masses measured in GeV. Charge conservation from Eq.~\eqref{eq::charge_con} implies that $Q_1 = -m_1r_\rho/m_2$ for $Q_2=1$, so one obtains Eq.~\eqref{eq::alpha_final}, which can be directly plugged into Eq.~\eqref{eq:Rlight}.

\bibliographystyle{my-h-physrev}
\bibliography{multiDM_extension}

\end{document}